\documentclass[12pt]{iopart}

\usepackage{graphicx}
\usepackage{tikz}
\usepackage{pgfplots}
\pgfplotsset{compat=1.5}
\usepackage{pgfplotstable}
\usetikzlibrary{calc}
\tikzset{mark options={mark size=5, line width=1pt}}

\begin{document}
%opening
%\title[Bright and versatile high rep. rate XUV beamline for coincidence imaging]{Bright and versatile high repetition rate extreme ultraviolet beamline for coincidence electron-ion imaging}
\title[ ]{Bright, polarization-tunable high repetition rate extreme ultraviolet beamline for coincidence electron-ion imaging}
\author{A. Comby$^{1}$,
E. Bloch$^{1}$, 
S. Beauvarlet$^{1}$,
D. Rajak$^{1}$,
S. Beaulieu$^{3}$,
D. Descamps$^{1}$,
A. Gonzalez$^{2}$,
F. Guichard$^{2}$,
S. Petit$^{1}$,
Y. Zaouter$^{2}$, 
V. Blanchet$^{1}$,
Y. Mairesse$^{1}$}

\address{$^1$ Universit\'e de Bordeaux - CNRS - CEA, CELIA, UMR5107, F33405 Talence, France}
\address{$^2$ Amplitude Laser Group, 33600 Pessac, France}
\address{$^3$ Department of Physical Chemistry, Fritz Haber Institute of the Max Planck Society, Faradayweg 4-6, 14195 Berlin}
\ead{yann.mairesse@u-bordeaux.fr}
\vspace{10pt}
\begin{indented}
\item[]February 2020
\end{indented}

\begin{abstract}
After decades of supremacy of the Titanium:Sapphire technology, Ytterbium-based high-order harmonic sources are emerging as an interesting alternative for experiments requiring high flux of ultrashort extreme ultraviolet (XUV) radiation. In this article we describe a versatile experimental setup delivering XUV photons in the 10-50 eV range. The use of cascaded high-harmonic generation enables us to reach 1.8 mW of average power at 18 eV. Several spectral focusing schemes are presented, to select either a single harmonic or group of high-harmonics and thus an attosecond pulse train. In the perspective of circular dichroism experiments, we produce highly elliptical XUV radiation using resonant elliptical high-harmonic generation, and circularly polarized XUV by bichromatic bicircular high-harmonic generation. As a proof of principle experiment, we focus the XUV beam in a coincidence electron-ion imaging spectrometer, where we measure the photoelectron momentum angular distributions of xenon monomers and dimers. 
\end{abstract}

\section{Introduction}
Since its discovery at the end of the 1980s \cite{mcpherson87,ferray88}, the production of coherent ultrashort extreme ultraviolet (XUV) radiation by high-order harmonic generation (HHG) has led to many applications in atomic, molecular and solid state physics, with temporal resolution down to the attosecond range \cite{lhuillier03,krausz09,lepine14,nisoli17}. These achievements were made possible thanks to the combination of fundamental research discoveries and the emergence of new laser technologies. The first high-order harmonic spectra were produced using 36 ps pulses at 1060 nm from a Nd:YAG laser \cite{ferray88}, and sub-ps pulses at 248 nm from a dye laser \cite{mcpherson87}. The developments of Ti:Sa lasers were a major breakthrough, and these sources are to date the most widely used in HHG experiments. Ti:Sa lasers combine the advantage of delivering short pulses (typically 25 fs) with high energy (1-100 mJ), at repetition rates in the 10 Hz - 10 kHz. These pulses can efficiently be converted into extreme ultraviolet by HHG up to the microjoule level \cite{hergott02,takahashi02}, and be used for non-linear XUV photoionization experiments \cite{takahashi02,tzallas03,takahashi13,senfftleben19}. The postcompression of laser pulses down to the sub-5 fs range in hollow core fibers \cite{nisoli96}, associated with the active stabilization of the carrier-envelop phase \cite{xu_route_1996}, enables the generation of single attosecond pulse \cite{hentschel01,chini14}. High energy radiation, reaching the water window (282-533 eV), can be produced by converting the pulses to the mid-infrared range using optical parametric amplifiers \cite{chen10,ren18}. 

Despite all these impressive achievements and the numerous applications, one important goal remained out of reach of the HHG sources based on Ti:Sa technology: the production of high repetition rate XUV pulses, beyond 100 kHz. Increasing the repetition rate of a source can have many benefits for applications which require a high average flux, but not necessarily a high XUV intensity. This is for instance the case of coincidence electron-ion experiments, in which the number of events per pulse is restricted to 1/10 to avoid false coincidences \cite{ullrich03}. High repetition rate XUV sources are also beneficial for time-resolved angle-resolved photoemission spectroscopy from surfaces, where a high number of photons per pulse produces space-charge, leading to detrimental shifts and broadening of the photoelectron spectra, but where the multidimensional detection scheme requires a high counting statistics, thus requiring a large number of pulses/second \cite{schonhense18,corder18,saule19}.

Ti:Sa systems suffer from the short upper state lifetime, their large quantum defect and the fact that they must be pumped with high brightness green lasers. This limits the efficiency, output average power, and prevents repetition rate scaling to drive strong field physics experiments. The technological breakthrough that made high repetition rate HHG possible is the advent of Yb-doped amplifiers. These lasers can reach higher average powers than Ti:Sa, up to the kW level for both bulk and fiber amplifiers \cite{russbueldt10,eidam10}. Moreover, Yb-doped fiber amplifiers (YDFA) can produce short pulses, down to typically 120 fs \cite{ruehl10,lavenu17}. On the other hand, the propagation in long fibers restricts the maximum energy per pulse that can be obtained, typically in the few \linebreak 100 $\mu$J range.

The first demonstration of HHG from YDFA sources was achieved in 2009 \cite{boullet09}. The low energy and long pulse duration impose tighter focusing conditions than experiments relying on Ti:Sa sources. This was initially thought to be detrimental, but the works of Rothhardt \textit{et al.} \cite{rothhardt14-2} and Heyl \textit{et al.} \cite{heyl16} demonstrated that the HHG efficiency remains the same whatever the focusing conditions, as long as the generating medium parameters (gas jet length and pressure) are properly scaled. The optimization of high repetition rate HHG thus led to the generation of XUV beams with average flux of several \linebreak 100 $\mu$W \cite{wang15,klas16}. Note that higher XUV fluxes, in the mW range, were reported based on a completely different technology, namely intra-cavity HHG at 77 MHz\cite{porat18-1}.

In this article we aim at describing an XUV beamline for coincidence electron-ion imaging of gas-phase photoionization. Our goal is to produce bright ultrashort radiation with controlled polarization state in the 10-50 eV range. We are particularly interested in the low energy range, in the outer valence ionization region of many molecular species, where molecular chirality can be probed with high sensitivity by photoelectron circular dichroism using circularly polarized radiation \cite{ritchie76,bowering01,nahon15}. After the general description of the beamline and its requirements, we discuss the optimization procedure of the high-harmonic source by cascaded harmonic generation. 
Section 3 describes different cases of spectral selection of the harmonic radiation, using metal foils and a normal incidence polarization-preserving monochromator. Next, we raise the issue of producing circularly polarized XUV radiation and present the results of two experiments, elliptical HHG and bicircular bichromatic HHG. Last, we present a coincidence electron-ion measurement of the photoionization of Xe atoms and dimers by high-order harmonics.

\section{Overview of the beamline}
\label{overview}
Our beamline is based on the BlastBeat laser system, which is made of two synchronized Tangerine Short Pulse Yb-doped fiber amplifiers seeded by the same oscillator (Amplitude). Each amplifier delivers a 50 W average power beam of 135 fs pulses centered at 1030 nm, with a repetition rate tunable between 166 kHz and 2 MHz. The highest energy per pulse is thus 300 $\mu$J at 166 kHz. The dual configuration aims at using one amplifier to generate spectrally-tunable pump pulses by frequency conversion, and the other one to produce XUV radiation as probe pulses through HHG. In this article we focus on the probe XUV beamline, using a single 50 W beam. 

\begin{figure}
\begin{center}
\includegraphics[width=\textwidth]{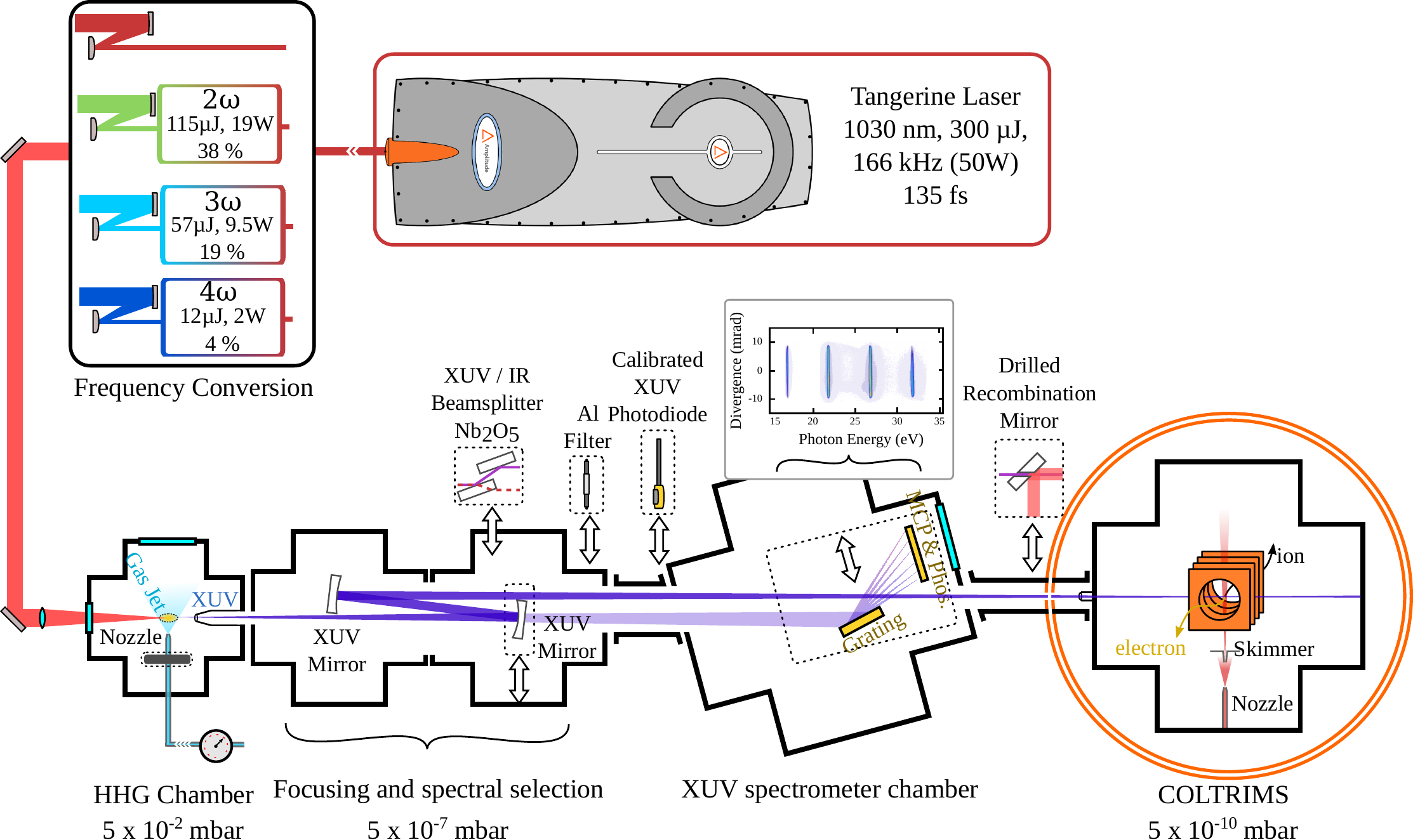}
\caption{Overview of the beamline. The beam from an YDFA is frequency upconverted before being focused into a vacuum chamber to produce high-order harmonics. A telescope made of two spherical XUV mirrors can be inserted to spectrally select some harmonics, and refocus them 3 m away into the interaction chamber of a COLTRIMS. The typical pressures in the different chambers are indicated at the bottom of the figure. }
\label{fig1}
\end{center}
\end{figure}

High-order harmonics can be directly produced from the output of the laser amplifier, or after frequency conversion. Our setup can produce the second, third and fourth harmonics of the laser, using non-linear crystals (for a detailed description see \cite{comby19}). The resulting beam parameters are listed in Table \ref{table_omega_L}.

\begin{table}[! h ]
\centering
\begin{tabular}{|c|c|c|c|c|}
  \hline
  Harmonic & 1 & 2 & 3 & 4\\
  \hline
  Laser frequency & $\omega_L$ & 2$\omega_L$  & 3$\omega_L$  & 4$\omega_L$ \\
  \hline
  Wavelength (nm) & 1030 & 515  & 343  & 257.5 \\
  \hline
  Photon energy (eV) & 1.2 & 2.4 & 3.6  & 4.8 \\
  \hline
  Average power (W) & 50 & 19 & 9.5  & 2 \\
  \hline
  Energy per pulse ($\mu$J) & 300 & 115 & 57  & 12 \\
  \hline
  Conversion efficiency & - & 38$\%$ &19$\%$ & 4$\%$ \\
  \hline
\end{tabular}
\caption{Characteristics of the frequency conversion of the laser at 166 kHz. }
\label{table_omega_L}
\end{table}

In order to control the spot size and to avoid non-linear and/or thermal effects in the vacuum chamber entrance window, each beam can be expanded by an all-reflective telescope, before being focused by a lens into a vacuum chamber pumped by a 600 m$^3$/h roots pump, where it interacts with a continuous gas jet. The beam magnification before focusing enables us to keep all optical elements outside the vacuum chamber, simplifying the alignment procedure and minimizing thermal effects on the optics and optical mounts, which are much more critical under vacuum. The nozzle position with respect to the laser focus is adjusted using motorized translation stages. A differential pumping hole (located 3 mm away from the laser focus, with an increasing diameter from 0.5 mm to 5 mm over a 8 cm length), ensures that the generated high order harmonics are not strongly reabsorbed by the relatively high pressure of the HHG chamber (typically $5 \times 10^{-2}$ mbar). 

A CCD camera monitors the plasma produced during the high-order harmonic generation process, enabling us to retrieve the absolute density profile of the gas jet\cite{comby18,horke17}.  The plasma light emission is calibrated by filling in the chamber with a static pressure of argon, and measuring the plasma light intensity as a function of gas density. The resulting calibration curve is used to determine the gas jet profile, giving access to the exact macroscopic HHG parameters (typically 350 $\mu$m medium length, $2.5 \times 10^{18}$ atoms/cm$^3$ density) with a remarkably simple scheme.

The driving laser and XUV beams enter a second vacuum chamber, pumped by a 500 l/s turbomolecular pump down to $5 \times 10^{-7}$ mbar, where different elements can be inserted. A spherical mirror (f = 60 cm) can be placed onto the beam path to collimate the XUV beam, which is then focused by a second XUV mirror (f = 300 cm) into the COLTRIMS (COLd Target Recoil Ion Momentum Spectrometer), located at the end of the beamline. The mirrors are placed into motorized mounts, and the first one is on a motorized translation stage to adjust the focus position in the COLTRIMS. The magnification of the refocusing telescope is 5. With a typical size of the high-harmonic source of 10 $\mu$m, the spot size of the XUV in the COLTRIMS is thus in the range of 50 $\mu$m. The mirrors can ensure a spectral selection of the high harmonics, but they also partly reflect the fundamental laser beam. Additional spectral filtering is thus necessary, and can be achieved by inserting thin metal foils (Al, In). To protect these foils from the laser power, fused silica plates with a Nb$_2$O$_5$ coating can be inserted under 20$^\circ$ grazing incidence. The couple of plates reflects 42$\%$ and 0.7$\%$ at 1030 nm beam in s and p polarizations respectively, and 60$\%$ and 0.7$\%$ at 515 nm. In the XUV range, from 20 to 45 eV, the two plates reflect between 20 and 50$\%$ in s polarization and between 10 and 30$\%$ in p \cite{comby19}. 

At the exit of the second vacuum chamber, a calibrated photodiode can be used to measure the absolute photon flux. This photodiode is placed behind two aluminum filters, which can be alternatively removed and whose transmission can thus be calibrated. The third vacuum chamber is devoted to XUV spectrum measurement. It contains a gold-coated grating with variable groove spacing (1200 mm$^{-1}$ average), which images the spatially-resolved spectrum on a set of chevron microchannel plates with a fast (P46) phosphor screen. The decay time of the phosphor (500 ns to reach 10$\%$ of the signal) is lower than the temporal spacing between two consecutive laser pulses (6 $\mu$s at 166 kHz), avoiding saturation of the screen. The grating is placed on a mechanical translation stage and can be removed to let the beam reach the COLTRIMS apparatus. The XUV beam then goes through the hole of a drilled mirror under 45$^\circ$, which can recombine the pump and the probe (XUV) beams.

Between the spectrometer and the COLTRIMS chambers, a differential pumping stage (3 mm hole, 300 L/s turbo pump) is placed. The background vacuum in the COLTRIMS target chamber is $5\times 10^{-10}$ mbar and is insensitive to the pressure in the HHG chamber. Two consecutive optical skimmers with diameters 9 and 6 mm are placed along the differential pumping stage. Their centering can be finely adjusted, to filter out stray XUV light. At the exit of the COLTRIMS, an inverted optical skimmer is placed, and the beam is reflected by a $45^\circ$ XUV mirror and directed into a beam dump. This avoids backscattering by the COLTRIMS exit window, which can induce significant signal on the electron detector. 

The target gas is introduced into the COLTRIMS by a 30 $\mu$m nozzle, which can be heated to introduce compounds that were initially solid or liquid, pushed by a carrier gas. The center of the supersonic gas jet is selected using a 200 $\mu$m diameter skimmer placed 9 mm away from the nozzle. The selected gas jet goes travels through the interaction chamber until a double stage beam dump. In the interaction chamber, the electron-ion \linebreak spectrometer is made of a stack of electrodes producing a homogeneous electric field (8.2 V.cm$^{-1}$) over a 29.1 cm flight tube on the electron side and 4.1 cm on the ion side. An homogeneous magnetic field (8.25 Gauss) generated by Helmholtz coils shapes the electron trajectories to ensure that all electrons reach the 80 mm diameter detector. The geomagnetic field is compensated using additional correction coils. The photoelectron signal is amplified with a stack of two microchannel plates (MCP) and detected with hexagonal (6 channels) delay lines, which provide the position and time-of-flight of the particles. The photoions are detected on the other side using MCPs and square (4 channels) delay lines. The ions and electrons can be detected in coincidence, and their momentum can be fully reconstructed in three dimensions as the detection is position- and time-of-flight-sensitive \cite{ullrich03}. 

\section{Cascaded high-harmonic generation}
Cascaded high-harmonic generation consists in producing high-order harmonics from low-order harmonics of the fundamental frequency. We have described in Section \ref{overview} the generation of second, third and fourth harmonics from the 1030 nm pulses. Using these harmonics to drive the HHG process has several advantages. 

\subsection{Advantages}
\textbf{Increased efficiency:} This is the most obvious advantage of cascaded HHG.  Many theoretical and experimental studies \cite{balcou92,colosimo08,shiner09,lai13,marceau17,wang15,popmintchev15} have established that the harmonic dipole moment $d_{HHG}$ scales with the driving laser wavelength $\lambda_L$ (and driving laser frequency $\omega_L$) as :
\begin{equation}
d_{HHG} \propto \lambda_L^{-\alpha} \equiv d_{HHG} \propto  \omega_L^{\alpha}
\label{loi_lambda}
\end{equation}
where $\alpha$ varies between 4 and 8 depending on the spectral range and generating conditions. Quadrupling the laser frequency may potentially increase the HHG conversion efficiency by a factor ranging from 256 to more than 65000. This gain will partly be mitigated by the limited conversion efficiency of the optical harmonic generation, but the situation should be beneficial in all cases.

\textbf{Increased spectral interval:} A second advantage of cascaded high-harmonic generation is the increased spectral separation between consecutive harmonics. High-order harmonics are separated by twice the laser frequency, i.e. 2.4 eV at 1030 nm, and 9.6 eV at 257 nm. Driving the HHG process with shorter wavelength thus makes the spectral selection of an isolated harmonic easier, using multilayer mirrors or metal foils. Furthermore, photoionization experiments from polyatomic molecules often produce congested photoelectron spectra, because each photon energy opens many ionization channels. The increased spectral interval between harmonics eases the assignment of such spectra. 

\textbf{Lower laser power:} A side benefit of the increased efficiency of the HHG process is that less laser power is necessary to produce the XUV radiation. Handling several tens of Watts of laser power can indeed be problematic. The laser beam can heat the XUV optics, inducing beam deformation, pointing drifts and even damaging them. This can be handled by using thick substrates with good thermal properties, such as Si, and cooling the optical mounts. More critically, many experiments use thin metal foils to remove the fundamental laser, and spectrally select the harmonics of interest. From our experience, a 150 nm self-supported aluminum foil melts as soon as it is exposed to more than 5 W/cm$^2$ of radiation. With 50 W of available laser power to drive HHG at 1030 nm, this sets a severe constraint on the set-up. The laser beam can be strongly attenuated by using XUV-laser beamsplitters, such as the SiO$_2$ plates covered with Nb$_2$O$_5$ described in Section \ref{overview}. However these plates are incompatible with the generation of circularly polarized high-order harmonics, since they do not preserve the polarization state when reflecting elliptical radiation. An appealing solution consists in producing high-harmonics from an annular beam, which can be blocked by a pinhole transmitting the XUV light. This solution was successfully applied to a 5 W beam \cite{klas18}, but producing an annular beam with several tens of W of average power could become a challenging task. Starting from a shorter wavelength driver, with low power, is thus technically interesting. 

\subsection{Drawbacks}
Despite all these benefits, cascaded high-harmonic generation also presents a number of drawbacks: 

\textbf{Lower cutoff:} The highest photon energy that can be produced in HHG is determined by the maximum kinetic energy an electron can gain in the driving laser field. This energy scales as the square of the driving wavelength. Consequently, cascaded HHG produces lower photon energies. Note that if the laser intensity is high enough to fully ionize the medium and produce significant HHG from the ions, this limitation can be overcome \cite{popmintchev15}.

\textbf{UV optics:} Handling high power femtosecond laser beams in the ultraviolet range is difficult because of linear and non-linear absorption effects in transmitting media. This issue can be partly circumvented by using all-reflective optics to focus the laser beam, but in any case it has to go through a vacuum window to reach the gas jet. The optical quality of this window becomes a critical parameter in experiments. We have tested several windows using our 2 W beam at 257 nm, and found out that only high quality CaF$_2$ could handle more than 8 hours of operation without showing the appearance of colored centers.

\subsection{Optimized harmonic photon flux}
\begin{figure}
\begin{center}
\includegraphics[width=\textwidth]{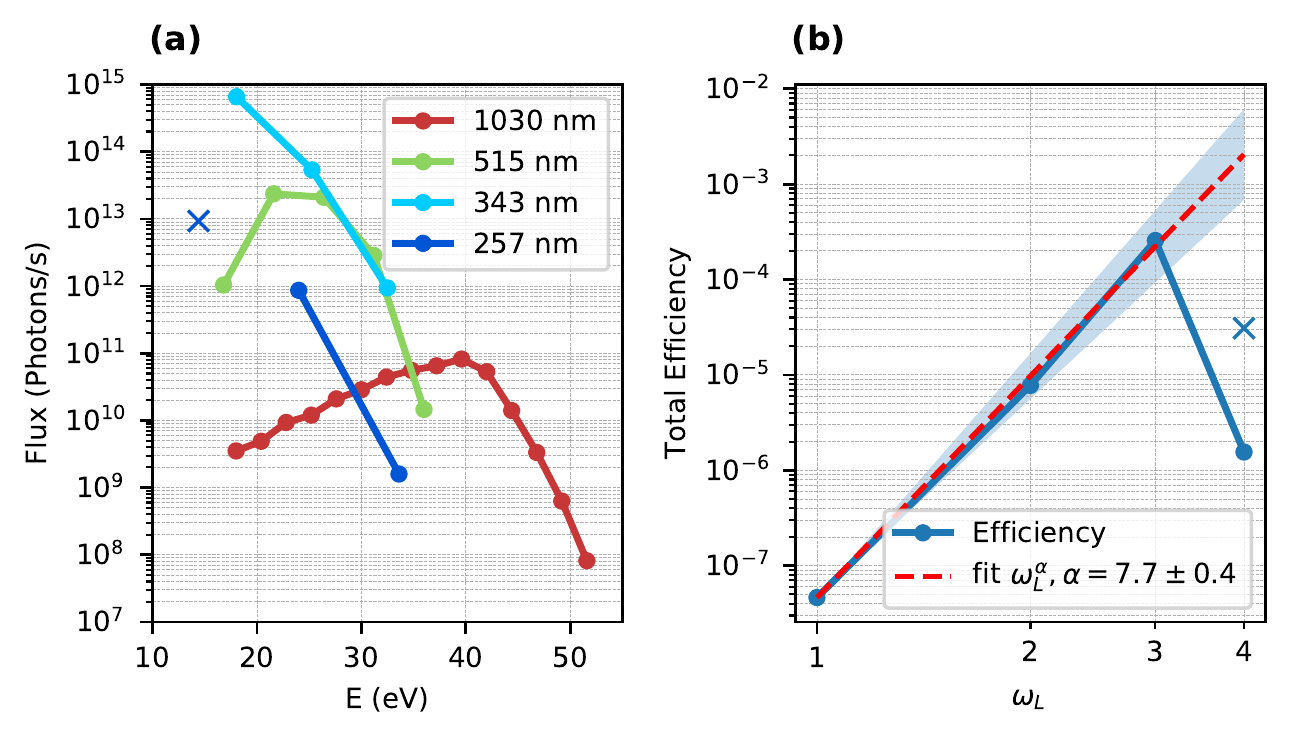}
\caption{Scaling of the cascaded HHG process in argon with driving wavelength. (a) Flux measurement and (b) efficiency with a power fit for the first three $\omega_L$. These measurements were presented in \cite{comby19}. The dark blue crosses on (a) and (b) correspond to new measurements of the optimized third harmonic generation from the $4\omega_L$, at 14.4 eV.}
\label{figscaling}
\end{center}
\end{figure}

In a previous work, we have quantitatively investigated the scaling of the HHG efficiency with driving laser wavelength \cite{comby19}. The results are summarized in Fig. \ref{figscaling} and Table \ref{tablescaling}

\begin{table}[t]
\centering
\begin{tabular}{| c || c | c | c | c |}
\hline
 $\omega_L$ & Photon energy (eV) & 25.2 (H21) & 39.6 (H33) & 44.4 (H37) \\
\cline{2-5}
1030 nm  & Photon flux ($ \times 10^{10}$ photons/s)  & 1.2$\pm$0.2 & 8.2$\pm$1.5 & 1.4$\pm$0.3 \\
\cline{2-5}
45 W & Power (nW) & 50$\pm$10 & 520$\pm$100 & 100$\pm$20 \\
\hline \hline

$2\omega_L$ & Photon energy (eV) & 21.6 (H9) & 26.4 (H11) & 31.2 (H13) \\
\cline{2-5}
515 nm & Photon flux ($ \times 10^{13}$ photons/s)  & 2.3$\pm$0.4 & 2.1$\pm$0.4 & 0.3$\pm$0.05 \\
\cline{2-5}
19 W & Power ($\mu$W) & 80$\pm$15 & 90$\pm$20 & 15$\pm$3 \\
\hline \hline

$3\omega_L$  & Photon energy (eV) & 18 (H5) & 25.2 (H7) & - \\
\cline{2-5}
343 nm & Photon flux ($ \times 10^{14}$ photons/s)  & 6.6$\pm$1.3 & 0.5$\pm$0.1 & - \\
\cline{2-5}
7.3 W & Power (mW) & 1.9$\pm$0.4 & 0.20$\pm$0.04 & - \\
\hline \hline

$4\omega_L$ & Photon energy (eV) & 14.4 (H3) & 24 (H5) & - \\
\cline{2-5}
257 nm & Photon flux ($ \times 10^{12}$ photons/s)  & 9$\pm$2 & 0.8$\pm$0.2 & - \\
\cline{2-5}
2 W & Power ($\mu$W) & 21 $\pm$ 4 & 3$\pm$1 & - \\
\hline

\end{tabular}
\caption{Measured XUV photon flux in optimized conditions obtained by cascaded HHG from a fundamental at  1,2,3 et 4 $\omega_L$. The most relevent photon energies are presented.}
  \label{tablescaling}
\end{table}

At $\omega_L$, the XUV photon flux reaches almost 10$^{11}$ photons/s near 35-40 eV (500 nW average power), from 45 W of laser power. Doubling the frequency of the generating pulse increases the XUV flux by two orders of magnitude, providing $2.1 \times 10^{13}$ photons/s (90 $\mu$W) at 26.4 eV starting from 19 W of 2$\omega_L$. The HHG process is even more efficient at 3$\omega_L$, producing $6.6 \times 10^{14}$ photons/s at 18 eV from 7.3 W of 3$\omega_L$. This raises the HHG source to the mW range, with 1.9 mW of average power at 18 eV. 
 
When the laser frequency is further increased to 4$\omega_L$, the benefit from the cascade does not seem to be that strong : the flux drops to $8 \times 10^{11}$ photons/s (3.3 $\mu$W) at 24 eV, \linebreak  from 2 W of 4$\omega_L$, which is still higher than the flux at $\omega_L$ but with a much lower scaling than observed at 2 and 3$\omega_L$. The low conversion efficiency from the 257 nm may seem surprising at first sight. However, one has to keep in mind that the ponderomotive energy at this wavelength is very low, such that the harmonic cutoff frequency at an intensity of $ 1.5 \times 10^{14}$  W.cm$^{-2}$ is expected to be around 17 eV, much below the energy of harmonic 5 (24 eV). We thus repeated the measurements by optimizing the generation and detection of harmonic 3 (14.4 eV). We obtained a flux of $9 \times 10^{12}$ photons/s (21 $\mu$W), from 0.7 W of 4$\omega_L$, in conditions in which harmonic 5 was hardly detected. As we will see in section 4, this configuration is quite interesting for applications in near-threshold molecular photochemistry. 

Finally, we investigated the wavelength dependence of the harmonic conversion efficiency, defined as the fraction of laser power converted in the XUV range. The maximum conversion efficiency is obtained by generating harmonics with the 3$\omega_L$ (343 nm), \linebreak  and reaches $2.6 \times 10^{-4}$ at 18 eV. The high value of this efficiency indicates that using rather long femtosecond pulses (135 fs) is not detrimental for the HHG conversion, at least in the low photon energy range. Our results show that the conversion efficiency scales as $\omega_L^\alpha$, with $\alpha = 7.7 \pm 0.4$. The scaling is more abrupt than the one predicted by resolving the time dependent Schr\"odinger equation \cite{colosimo08} ($\lambda^{-5.5} \equiv \omega_L^{5.5}$) . We interprete this deviation as the result of the strong ionization of the medium, which perturbs less shorter driving wavelengths, as established by Popmintchev \textit{et al.} \cite{popmintchev15}.

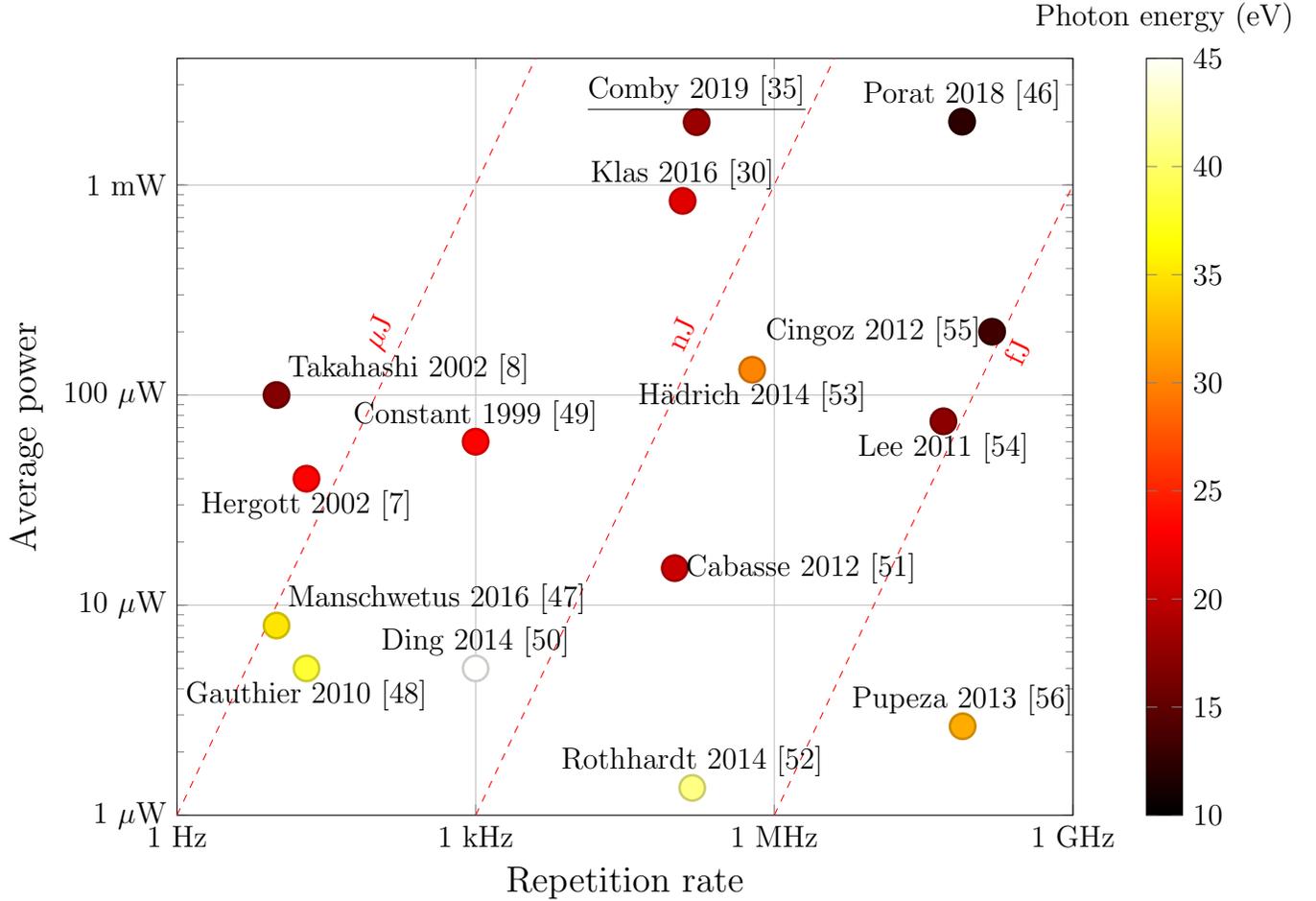
\begin{figure}
\centering
\begin{tikzpicture}
\begin{loglogaxis}[grid=major,width=14cm,colormap/%jet,colorbar,xmin=1e0,xmax=1e9,ymin=1e-15,ymax=1e-4,
hot2,colorbar,xmin=1e0,xmax=1e9,ymin=1e-6,ymax=4e-3,
		%point meta min=1,
		%point meta max=1.8,
		point meta min=10,
		point meta max=45,
		xlabel={\large{Repetition rate}},
		ylabel={\large{Average power}},
		xtick={1,1e3,1e6,1e9},
		xticklabels={{1 Hz},{1 kHz}, {1 MHz}, {1 GHz}},
		ytick={1e-6,1e-5,1e-4,1e-3},
		yticklabels={{1 $\mu$W},{10 $\mu$W}, {100 $\mu$W}, {1 mW}},
		colorbar style={title={Photon energy (eV)},
		ytick={10,15,20,25,30,35,40,45},
		yticklabels={10,15,20,25,30,35,40,45}}
		%yticklabel=$10^{\pgfmathparse{\tick}\pgfmathprintnumber\pgfmathresult}$}
		%colorbar style={} %,ymin=1,ymax=100}
		%colorbar style={ylabel={Energie de photon (eV)},zmode=log,ymin=1,ymax=100,ytick={1,10,100},yticklabel=$10^{\pgfmathprintnumber{\tick}}$}%yticklabels={1,10,100}
		]
    \addplot[
        scatter, mark=*, only marks, 
        scatter src=explicit,
        nodes near coords*={\Label}, 
        visualization depends on={value \thisrow{label} \as \Label}, %<- added value
        visualization depends on={value \thisrow{anchor}\as\myanchor},
   	 every node near coord/.append style={anchor=\myanchor},
    %] table [x=Hz,y=E,meta expr=log10(\thisrow{PE})] {
    ] table [x=Hz,y expr=\thisrow{E}*\thisrow{Hz},meta expr=(\thisrow{PE})] {
Hz	E	PE	label anchor
10	1e-05	16.7	 { Takahashi 2002 \cite{takahashi02}}	{south west}
%10	1e-06	26	\cite{takahashi_attosecond_2013}	{south east}
10	8e-07	35	{Manschwetus 2016 \cite{manschwetus16}}	{south west}
%10	1e-08	38	\cite{cassou_enhanced_2014}	east
%10	6e-09	58	\cite{cassou_enhanced_2014}	north
%10	7e-12	450	\cite{chen_bright_2010}	east
%2	4e-13	65	\cite{mcpherson_studies_1987}	west
%20	1.6e-11	1000	\cite{popmintchev_bright_2012}	south
20	2.5e-07	38	{Gauthier 2010 \cite{gauthier10}}	{north}
20	2e-06	23.2	 { Hergott 2002 \cite{hergott02}} 	north
1e3	6e-08	23.2	 {Constant 1999 \cite{constant99}} {south}
%1e3	1e-08	29.4	\cite{willner_coherent_2011}	south
1e3	5e-09	45	{Ding 2014 \cite{ding14}}	south
%1e3	5e-10	170	\cite{ding_high_2014}	east
%1e3	3e-10	98	\cite{lambert_optimized_2009}	north
120e3	7e-09	21.6	 {Klas 2016 \cite{klas16}}	{south}
%50e3	2e-09	22.3	 {Wang 2015 \cite{wang15}}	east
100e3	1.5e-10	20.4 {Cabasse 2012 \cite{cabasse12}}	west
150e3	9e-12	40.8 {Rothhardt 2014	\cite{rothhardt14}}	south
%100e3	6e-13	120	\cite{rothhardt_53_2014}	east
%100e3	1e-13	23.2	 {Heyl 2012 \cite{heyl12}}	east
%100e3	7e-14	38	{Lindner 2003 \cite{lindner03}}	north
600e3	2.2e-10	30	{H{\"a}drich 2014 \cite{hadrich14}}	north
%1e7	5e-11	8.3	{Ozawa 2015 \cite{ozawa15}}	{south}
%1e7	5.7e-12	27.7	 {Hadrich 2015 \cite{hadrich15}}	{north east}
%2.4e7	7.5e-17	23	{Emaury 2015 \cite{emaury2015}}	east
5e7	1.5e-12	17.2	 {Lee 2011 \cite{lee11}}	{north}
1.54e8	1.3e-12	13.3	 {Cingoz 2012 \cite{cingoz12}}	east
%7.8e7	1e-19	108	\cite{pupeza_compact_2013}	south
7.8e7	3.4e-14	32	{Pupeza 2013 \cite{pupeza13}}	south
%2.5e8	1.4e-14	20	{Carstens 2016 \cite{carstens16}}	{north east}
%2.08e7	4.8e-16	18	{Vernaleken 2011 \cite{vernaleken11}}	east
%1e8	1.6e-17	12	{Gohle 2005 \cite{gohle05}}	{south west}
7.70e7	2.6e-11	12.3 {Porat 2018	\cite{porat18}}	south
%166e3	1.2e-08	18	\cite{comby_cascaded_2019}	{south west}
%166e3	1.2e-09	25.2	\cite{comby_cascaded_2019}	west
%166e3	3.13e-12	39.6	\cite{comby_cascaded_2019}	east
166e3	1.2e-08	18	\underline{{Comby 2019 \cite{comby19}}} {south}
%166e3	1.2e-09	25.2	\textcolor{red}{[Comby 19]} 	west
%166e3	3.13e-12	39.6	\textcolor{red}{[Comby 19]} 	east
    };
    %\addplot[red,domain=1:1e9,samples=3, dashed] {1e-3*x} node[near start,above,sloped]{mJ};
    \addplot[red,domain=1:1e9,samples=3, dashed] {1e-6*x} node[near start,above,sloped]{$\mu$J};
	\addplot[red,domain=1000:1e12,samples=3, dashed] {1e-9*x} node[near start,above,sloped]{nJ};
	\addplot[red,domain=1e6:1e15,samples=3, dashed] {1e-12*x} node[near start,below,sloped]{fJ};
\end{loglogaxis}
\end{tikzpicture}
\caption{State of the art of high-order harmonic sources. The fundamental lasers used from 10 Hz to 1 kHz are Ti:Sa, from 100 kHz to 1 MHz are TDFA, and above 1 MHz are enhancement cavities. }
\label{bilan_flux}
\end{figure}

Our results can be compared with the recent works on high repetition rate HHG. Figure \ref{bilan_flux} shows the state of the art of high-order harmonic sources emitting \linebreak in the 10-100 eV range in terms of mean power at specific photon energy and repetition rate. This figure is inspired from the one presented in \cite{heyl17}. Klas \textit{et al.} \cite{klas16} measured a conversion efficiency of $7 \times 10^{-6}$ from a 120 W - 300 fs YDFA, and generated 0.83 mW at 21.3 eV in krypton after a post-compression stage. Porat \textit{et al.} \cite{porat18} reached a conversion efficiency of $2.5 \times 10^{-5}$ by generating harmonics in xenon using an enhancement cavity with a 80 W - 120 fs YDFA, producing 2 mW at 12.7 eV and 0.9 mW at 19.7 eV. These two impressive results are based on optical schemes which require a high level of expertise, and use costly gases. The overall efficiency of our cascaded HHG process, given by the ratio between the 1.9 mW of XUV energy produced at \linebreak 18 eV by the initial 50 W of fundamental laser power, is $(3.8  \pm 0.8) \times 10^{-5}$. The cascaded harmonic generation thus enables a clear global increase in efficiency. Furthermore the optical setup is a simple non-linear optics scheme, followed by a conventional HHG setup, ensuring a high robustness for long campaigns of experimental measurements, as we will show in section 4 and 6.
%\begin{figure}
%\begin{center}
%\includegraphics[width=\textwidth]{FigStateOfTheArt.png}
%\caption{State of the art of high-order harmonic sources in the 10-100 eV photon energy range. The pulse energy is shown as a function of the repetition rate. The references below 10 kHz used Ti:Sa lasers as primary sources. Above 100 kHz, the main technology is YDFA. \textbf{Check for higher references; remove high photon energy ones.} The color of each dot codes the photon energy of the corresponding source.  }
%\label{fig1}
%\end{center}
%\end{figure}

\section{Spectral selection and refocusing}
The fluxes presented in Fig. \ref{bilan_flux} correspond to the number of photons generated by the high-harmonic sources. However for most applications, it is necessary to filter out the generating laser beam, and often perform a spectral selection of a given harmonic or a bunch of high harmonics for attosecond spectroscopy.  We present several possibilities that we have implemented on our beamline, for the selection of a single harmonic of low energy, the broadband selection and refocusing of harmonics between 15 and 25 eV, and the selection and refocusing of a single harmonic line at 35 eV. 

\subsection{Low energy monochromatization}
We have seen that the high-order harmonic generation process starting from the fourth harmonic (257 nm) of the laser was not very efficient, because of the low ponderomotive energy of the electrons at this short wavelength, which imposes a low cutoff of the high-harmonic spectrum. Nevertheless, the production of the third harmonic of the $4\omega_L$ can be achieved with a rather good efficiency, reaching $3 \times 10^{-5}$ in argon at a intensity of $\sim 2 \times 10^{13}$ W/cm$^2$. The photon energy of this harmonic, 14.4 eV, can initiate photochemical processes, mimicking the absorption of extreme ultraviolet radiation by the upper atmospheres of planets. Such photochemical processes are generally investigated with discharge lamps, which are bright (10$^{14}$ photons/s/cm$^2$), but do not cover the 14-15 eV range \cite{tigrine16}. Full tunability is offered by synchrotron radiation, which delivers up to (10$^{12-14}$ photons/s/cm$^2$) \cite{nahon13}, but with a restricted access which limits the possibilities of systematic investigations. It was recently demonstrated that high-order harmonics from a Ti:Sa could be bright enough to trigger photochemical processes over a few hours of irradiation \cite{bourgalais19}. Cascaded high-harmonic generation could provide a brighter source for such investigations. 

A monochromatic source at 14.4 eV can be easily obtained by selecting the third harmonic of the $4\omega_L$ of the YDFA. A 150 nm thick indium foil has a measured transmission of $15 \%$ around 14.4 eV, and completely blocks the fundamental radiation at 257 nm. It can thus be used to monochromatize the XUV light. The cascaded scheme is very advantageous, because only 0.7 W of laser power are used to drive the harmonic generation. However, indium has a very low melting temperature, and we found out that even at this low power, the laser beam has to be attenuated before hitting the foil. A single SiO$_2$ plate under $30^\circ$ grazing incidence was inserted between the source and the foil, reflecting only $\sim 20 \%$ of the s polarized 257 nm. We measured the photon flux after the plate and the foil, and obtained $4 \times 10^{11}$ photons/s. This beam can directly be used for photochemical investigations which do not require refocusing, such as the astrochemistry experiments described above.

\subsection{Broadband spectral selection and refocusing -- 15-25 eV}
\begin{figure}
\begin{center}
\includegraphics[width=\textwidth]{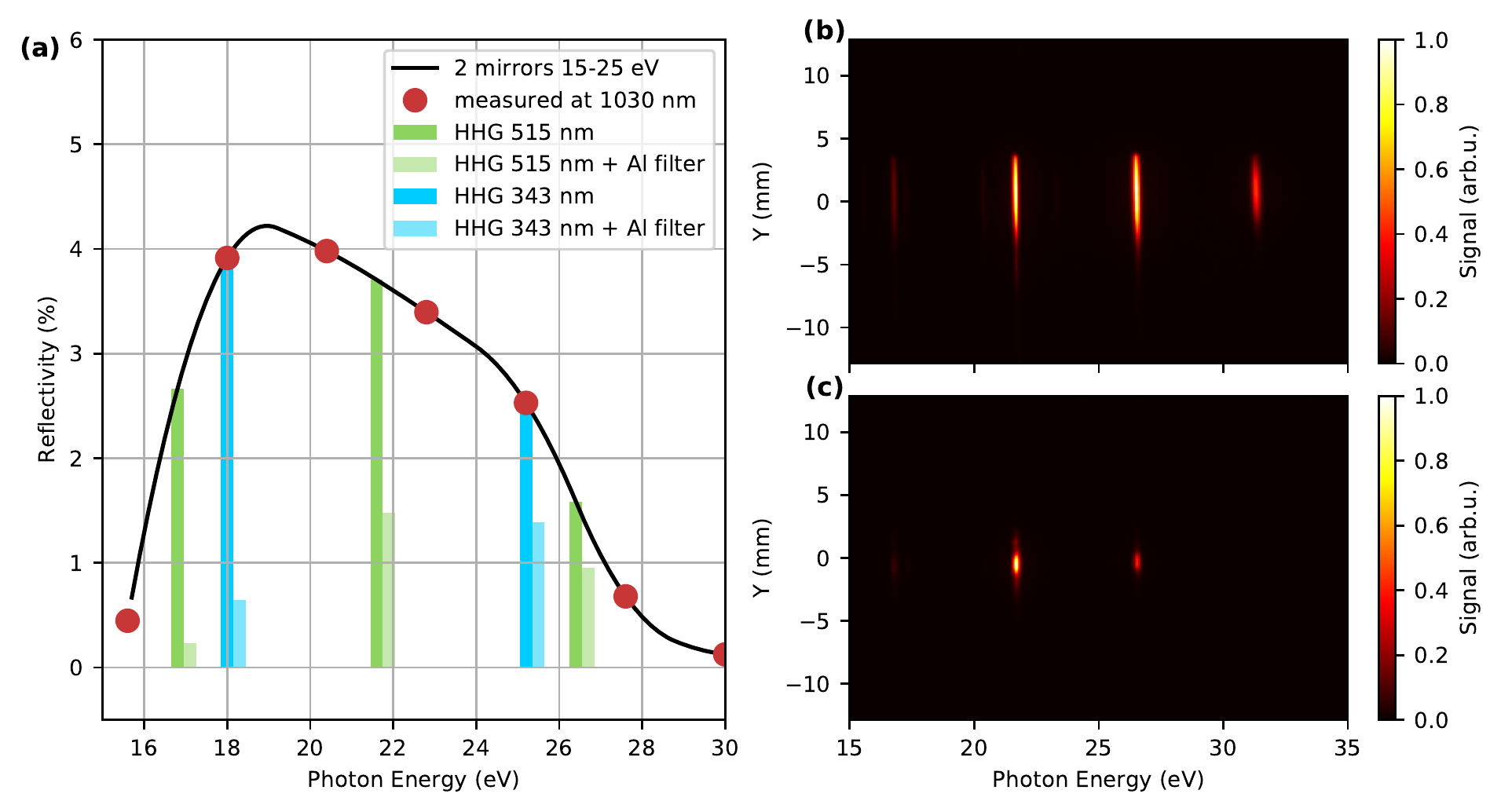}
\caption{Broadband spectral selection of high-harmonics by SiC mirrors at $0^\circ$ incidence. (a) Measured reflectivity of two mirrors (dots), interpolated reflectivity at the energy of the harmonics of the 515 nm and 343 nm, with and without 150 nm filter. Spatially resolved HHG spectrum before (b) and after (c) the two SiC spherical mirrors, showing both the spectral selection and focusing.}
\label{figbroadband}
\end{center}
\end{figure}

In the 15-25 eV spectral range, which is of interest for near-threshold photoelectron spectroscopy, it is extremely difficult to produce highly reflective XUV multilayer mirrors with a low enough bandwidth to isolate one harmonic in the HHG spectrum. However, there are many applications where using several harmonics can be important, e.g. for attosecond spectroscopy. Our beamline is thus equipped with a set of two SiC-coated spherical mirrors under quasi normal incidence, which image the harmonic source into the COLTRIMS chamber. In Fig. \ref{figbroadband} (a), we present the measured reflectivity of the pair of mirrors. The measurements were taken using high-harmonics from the fundamental 1030 nm laser beam, to maximize the number of sampled photon energies. The black line is a polynomial interpolation of the experimental data. The reflectivity shows a rather broad peak, above $2 \%$ between 17 and 25 eV. This scheme can be used to select a few harmonics of the $2\omega_L$ (515 nm) or $3\omega_L$ (343 nm) beams. The photon energies of these harmonics are shown in Fig. \ref{figbroadband}(a).  An aluminum foil can be added after the two mirrors to eliminate the fundamental radiation, and we plot the resulting transmission in Fig. \ref{figbroadband} (a). Figure \ref{figbroadband}(b-c) shows the spatially resolved harmonic spectra produced by the $2\omega_L$, and detected directly (b) or after being reflected by the pair of SiC mirrors (c). Two harmonics dominate the selected spectrum, with photon energies 21.6 eV \linebreak and 24 eV. 

The reflectivity of the SiC mirrors is much lower than what could be achieved using a grazing incidence toroidal mirror. However such optics can strongly modify the polarization state of the radiation, such that their use in circular dichroism experiments requires extreme caution \cite{barreau18}. The advantage of our scheme is the exclusive use of normal-incidence optics, which preserve the polarization state of the HHG. Despite the losses induced by these mirrors, our beamline delivers typically between $5 \times 10^{11}$ and $5 \times 10^{12}$ photons/s around 20 eV on target, starting from typically $ 10^{13}-10^{14}$ photons/s at the source (see Table \ref{bilan_flux}), which is more than enough for many applications.

\subsection{Narrowband spectral selection and refocusing -- 35 eV}
\begin{figure}
\begin{center}
\includegraphics[width=\textwidth]{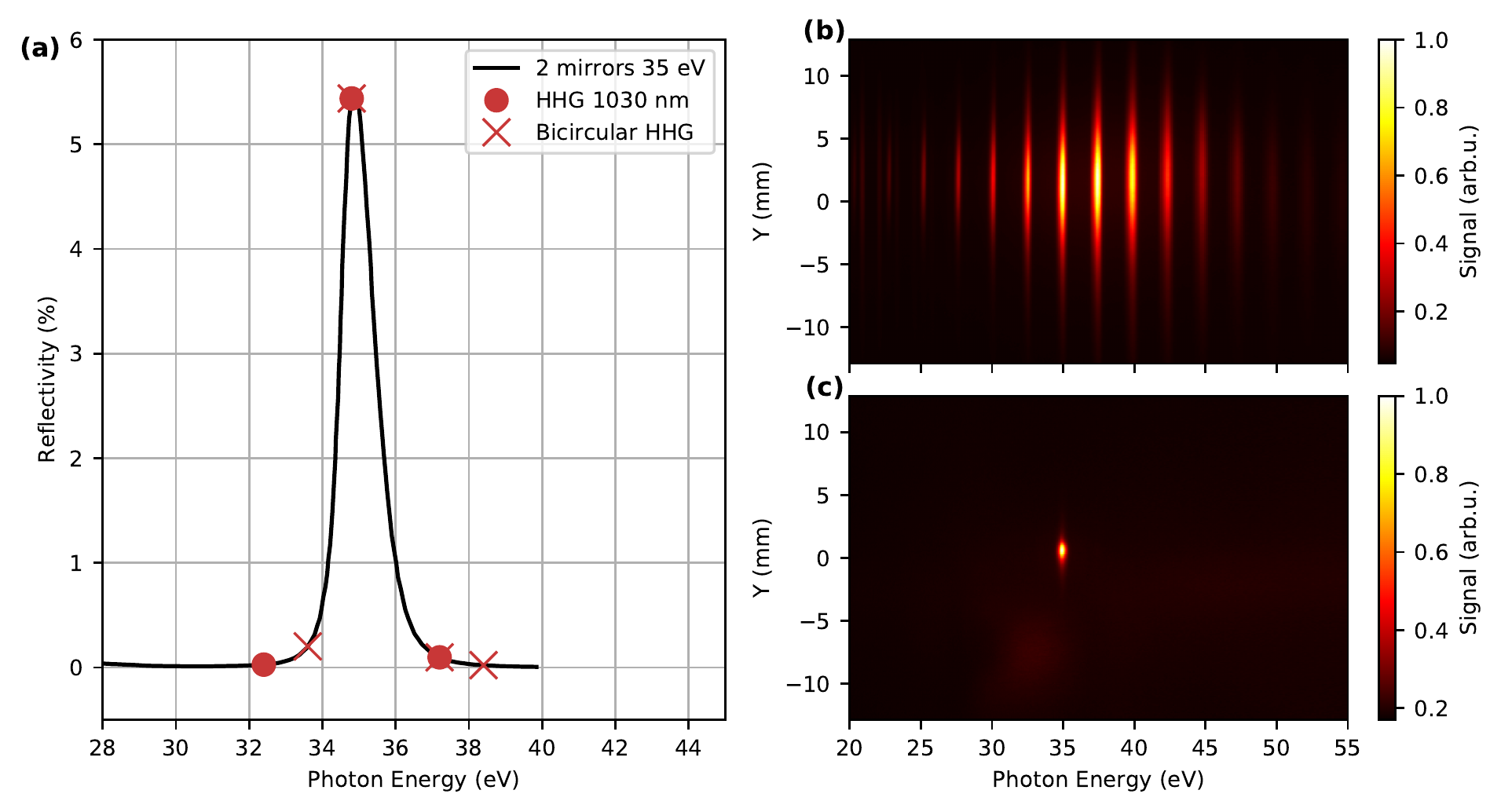}
\caption{Spectral selection of 35 eV high-harmonic by multilayer mirrors. (a) Calibrated reflectivity of a set of two 35 eV mirror at $0^\circ$ incidence. We place on this reflectivity curve the position of HHG generated from linearly polarized 1030 nm and from 515 + 1030 nm bicircular HHG. Spatially resolved HHG spectrum before (b) and after (c) multilayer mirrors, showing both the spectral selection and focusing.}
\label{fig35eV}
\end{center}
\end{figure}

The monochromatic source we have produced by isolating the third harmonic of the $4\omega_L$ (257 nm) at 14.4 eV is interesting for various purposes, but it is not refocused and thus cannot be used in the COLTRIMS. We have thus developed an alternative solution to obtain monochromatic XUV radiation, using highly selective multilayer mirrors. The reflectivity of these mirrors peaks at 35 eV, with a 1.5 eV full width at half maximum (FWHM) bandwidth (Fig. \ref{fig35eV}(a)). They can thus isolate harmonic 29 (34.9 eV) of the fundamental 1030 nm laser. The spatially resolved harmonic spectra obtained without and with two spherical multilayer mirrors are shown in Fig. \ref{fig35eV}(b-c). The reflectivity of the mirror pair is above $5\%$, leading to a flux around $1 \times 10^{9}$ photons/s on target. 

As we will show in the next section, these mirrors can also be used to select a circularly polarized high-harmonic produced by a combination of counter-rotating bicircular bichromatic fields \cite{milosevic00,kfir16}.

\section{Generation of circularly polarized high-order harmonics}
The generation of circularly polarized high-order harmonics has become an important topic in the past few years. Circularly polarized XUV radiation is a very interesting tool to investigate molecular chirality \cite{nahon15} or magnetic materials \cite{vanderlaan14}. Specific synchrotron beamlines are fully dedicated to XUV circular dichroism experiments. The femtosecond and attosecond temporal resolutions offered by high-order harmonic sources carry the promise to extend these dichroism experiments to the ultrafast regime. However generating circularly polarized high-harmonics is not straightforward, because the HHG process has a natural preference for linearly polarized laser fields, decaying exponentially with laser ellipticity. One solution to this issue is to introduce polarization-shaping elements onto the XUV beamline. Typically, a set of four mirrors under well-chosen incidence can play the role of an XUV quarter-waveplate, producing radiation with high degrees of ellipticity \cite{vodungbo11,willems15,siegrist19}. An alternative elegant solution consists in producing two collinear phase-locked high-order harmonic sources, with orthogonal polarizations. The polarization state of the resulting XUV radiation can be tuned by adjusting the delay between the two sources, which is analogous to changing the thickness of a birefringent crystal in the optical domain \cite{azoury19}. While very versatile, this solution is quite demanding in terms of interferometric stability. In the following, we present two schemes in which the HHG process in a single source is manipulated to directly produce circularly polarized high-harmonics. 

\subsection{Resonant high-harmonic generation}

One solution to produce highly elliptical XUV radiation directly from the HHG process is to perform resonant-HHG \cite{ferre15-1}. It was shown that high-harmonics generated in SF$_6$ molecules by a laser field with moderate ellipticity ($20\%$) could be highly elliptical (80$\%$), if the harmonics were emitted in an energy range where a resonance was at play in the generating process. This method being very straightfoward, we tried to extend it to the case of HHG from YDFA lasers. Unfortunately, we did not manage to produce any significant HHG signal from SF$_6$ molecules using 1030 nm or 515 nm pulses. We did not observe any significant plasma emission either. This could be surprising, because SF$_6$ has the same ionization potential as Ar, and should at first sight be ionized at the same laser intensity. However, the molecular orbitals of SF$_6$ have many nodes, which is an obstacle to efficient strong-field ionization \cite{muth-bohm00,tong02}. The previous HHG experiments in this molecule were using 800 nm and 400 nm pulses. These wavelengths are close to resonance with the interval between the lowest cationic states of the molecule, which could result in a strong enhancement of the ionization probability \cite{ferre15}. The 1030 nm and 515 nm fields may be too far from this ionic resonance to enable strong ionization of the molecules, preventing them to efficiently generate high-order harmonics. Other targets should thus be chosen to achieve highly elliptical XUV generation by resonant HHG -- for instance rare gases below their ionization threshold \cite{ferre15,antoine97-2}.

\begin{figure}[h]
\begin{center}
\includegraphics[width=\textwidth]{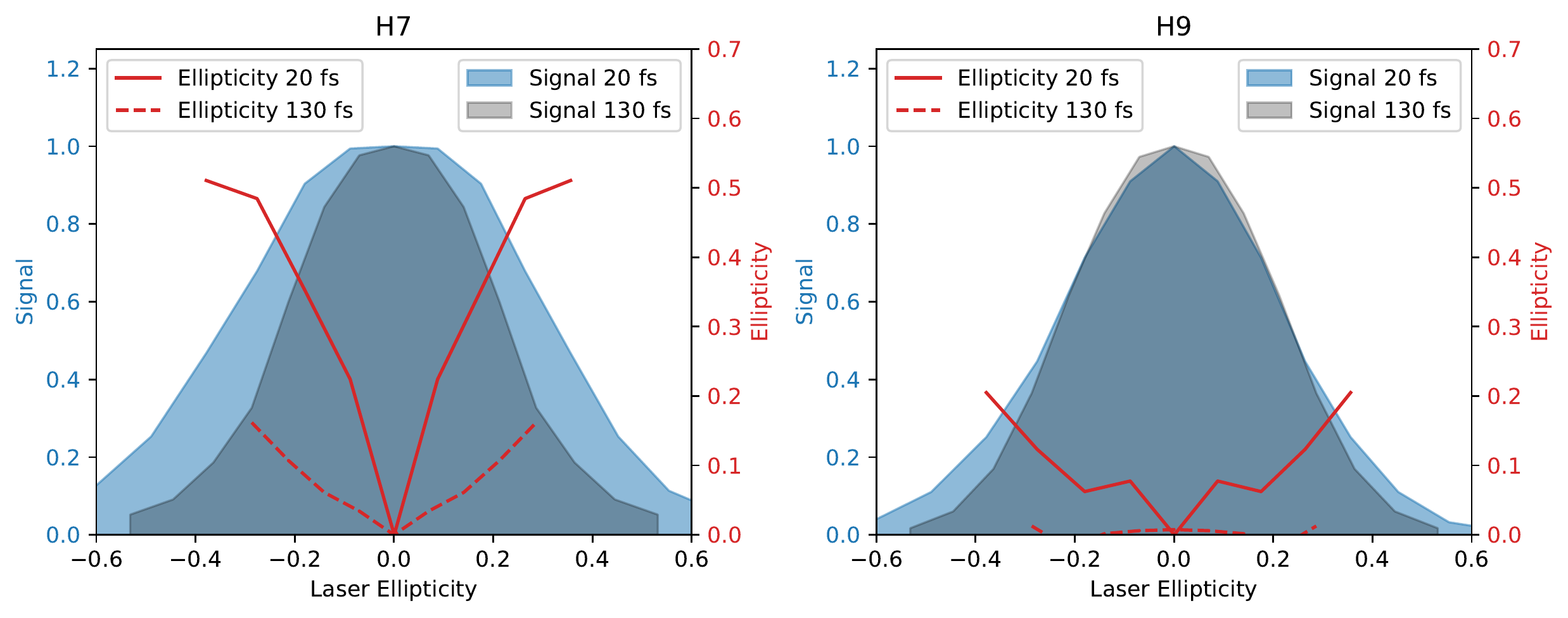}
\caption{Elliptical high-harmonic generation driven by 515 nm pulses in argon. Intensity (shaded area) and maximum ellipticity extracted from Malus' law (red line) of harmonic 7 (left) and 9 (right), using 130 fs pulses or postcompressed 20 fs pulses. }
\label{figresonant}
\end{center}
\end{figure}

In argon, the generation of highly elliptical XUV radiation was achieved by producing the fifth harmonic of a 400 nm laser beam. The photon energy of this harmonic is 15.5 eV, lying just below the ionization threshold of Argon (15.76 eV). The resonant excitation of the Rydberg states lying in this spectral region was found to explain the high ellipticity of the XUV radiation \cite{ferre15}. In order to study the possibility to produce highly elliptical XUV radiation with our YDFA source, we performed a ellipticity resolved-HHG experiment. We controlled the ellipticity of the laser by rotating a half waveplate in front of a fixed quarter waveplate. This allows to keep the main axis of the fundamental  polarization ellipse unchanged. In order to measure the ellipticity of the XUV, we built an XUV polarizer made of 4 unprotected gold mirrors under 20$^\circ$ grazing incidence. This polarizer has a measured extinction ratio between s and p polarisation above 20 in the 15-30 eV range. It can be rotated under vacuum to provide Malus' law curves, from which the polarization direction and ellipticity of the radiation can be extracted, under the assumption that the light is fully polarized \cite{antoine97-2}. We have to keep in mind that this assumption is not always valid, and that in the end only the measurement of a dichroic signal \cite{veyrinas16,barreau18} or the use of an XUV quarter wave plate \cite{nahon04} can provide the accurate value of the ellipticity, the current method only providing an upper bound $\epsilon_{max}$. Figure \ref{figresonant}(a) shows the polarimetry measurements of harmonics 7 and 9 of the $2\omega_L$ (515 nm). The laser ellipticity was scanned from $-55 \%$ to $+55 \%$, and the polarizer was rotated from -200$^\circ$  to 200$^\circ$. 
When the laser ellipticity increases from 0 to $40 \%$, the harmonic signal decreases by one of magnitude. The signal decays exponentially with laser ellipticity $\epsilon_L$ \cite{budil93}, following $I_q(\epsilon_L)\propto e^{-\beta_q\epsilon_L^2}$, where $\beta_q$ is the decay rate for harmonic $q$. Gaussian fits of the data presented in Fig. \ref{figresonant}(a) give $\beta_7=13.3$ and $\beta_9=11.6$. These rather low values characteristic of a low intensity, short wavelength HHG process. They are very similar to those obtained by Budil \textit{et al.} for harmonic 9-13 of 600 nm pulses generated in Xenon. The harmonic maximum ellipticity extracted from the Malus law increases quasi linearly with $\epsilon_L$, but remains low, reaching only 15$\%$ at $\epsilon_L=30\%$ for H7, and remaining quasi null for H9. This is characteristic of non-resonant HHG. Indeed, H7 and H9 correspond to photon energies of 16.8 and 21.6 eV, at which the absorption/emission spectrum of argon atoms do not present any particular resonant feature. In these conditions, the high-harmonic ellipticity is known to be lower than that of the driving laser pulse \cite{antoine97-2}. 

We repeated this experiment with shorter laser pulses, bringing the 515 nm pulses from 130 nm down to 20 fs by postcompression in a hollow core fiber. The details of this scheme will be described elsewhere. This postcompression broadens the laser spectral width from 18 meV FWHM to 260 meV FWHM. The polarimetry measurements of H7 and H9 from the 20 fs - 515 nm pulses are presented in Fig. \ref{figresonant}. While the signal decay rate is similar to the long driver case for H9 ($\beta_9=10.6$), H7 shows a much lower decay rate $\beta_7=6.4$. Furthermore, the harmonic maximum ellipticity reaches 50$\%$ at $\epsilon_L=30\%$. These results are quite intriguing : Why would the harmonic decay with ellipticity be sensitive to pulse duration? How could highly elliptical harmonics be generated in the absence of a resonance ? We believe that this is due to the AC-Stark shift of the Rydberg states by the strong generating field. At an intensity of $1 \times 10^{14}$  W.cm$^{-2}$, the ponderomotive potential is  $U_p$= 2.5 eV. This shift is a good approximation of the AC-Stark shift of the ionization potential of the atom and high lying Rydberg states. Harmonic 7, at 16.8 eV, can thus be in the vicinity of Stark-shifted Rydberg states. The broad bandwidth of the pulses, achieved thanks to the postcompression, increases the probability to hit a resonance and to produce highly elliptical XUV light. This situation, in which the influence of below-threshold states is visible above the ionization threshold, is similar to the one observed in high-harmonic generation from aligned molecules \cite{soifer10}. It constitutes an interesting manifestation of dynamical Stark-shifts in HHG. However, since the process strongly depends on the laser intensity, we expect that the polarization state produced will vary in time, leading to a partial depolarization of the XUV light. This scheme may thus not be optimal for future circular dichroism experiments.

\subsection{Bicircular bichromatic HHG}
Another solution to produce circularly polarized high-harmonics is to use tailored laser fields. Combining a circularly polarized fundamental pulse and its counter-rotating second harmonic creates an electric field with a clover structure, which can efficiently generate high-harmonics. The selection rules of this scheme are such that harmonics $ 3 n \pm 1$, where $n$ is an integer, are produced \cite{milosevic00}. Interestingly, since the fundamental and second harmonic beams are circularly polarized, the consecutive harmonics are also circularly polarized, with alternating helicity \cite{fleischer14}. This scheme is particularly appealing for circular dichroism measurements based on absorption, such as XMCD, because opposite helicities are probed simultaneously at neighbouring photon energies. On the other hand, in photoionization of polyatomic molecules, the congested nature of the photoelectron spectra can lead to an overlap of the signals from the counter-rotating harmonics, leading to potential canceling of circular dichroism effects, as well as to huge difficulties in energy assignment of the spectroscopic structures. For such applications, it is necessary to select one harmonic only. This can actually be performed with the monochromator presented in the previous section. A bicircular bichromatic HHG with 1030 nm and 515 nm fields produces harmonic 28 (33.7 eV) and 29 (34.9 eV), with opposite circular polarization. The monochromator should select the latter, the former being 20 times weaker (see Fig. \ref{fig35eV}).

\begin{figure}[h]
\begin{center}
\includegraphics[width=\textwidth]{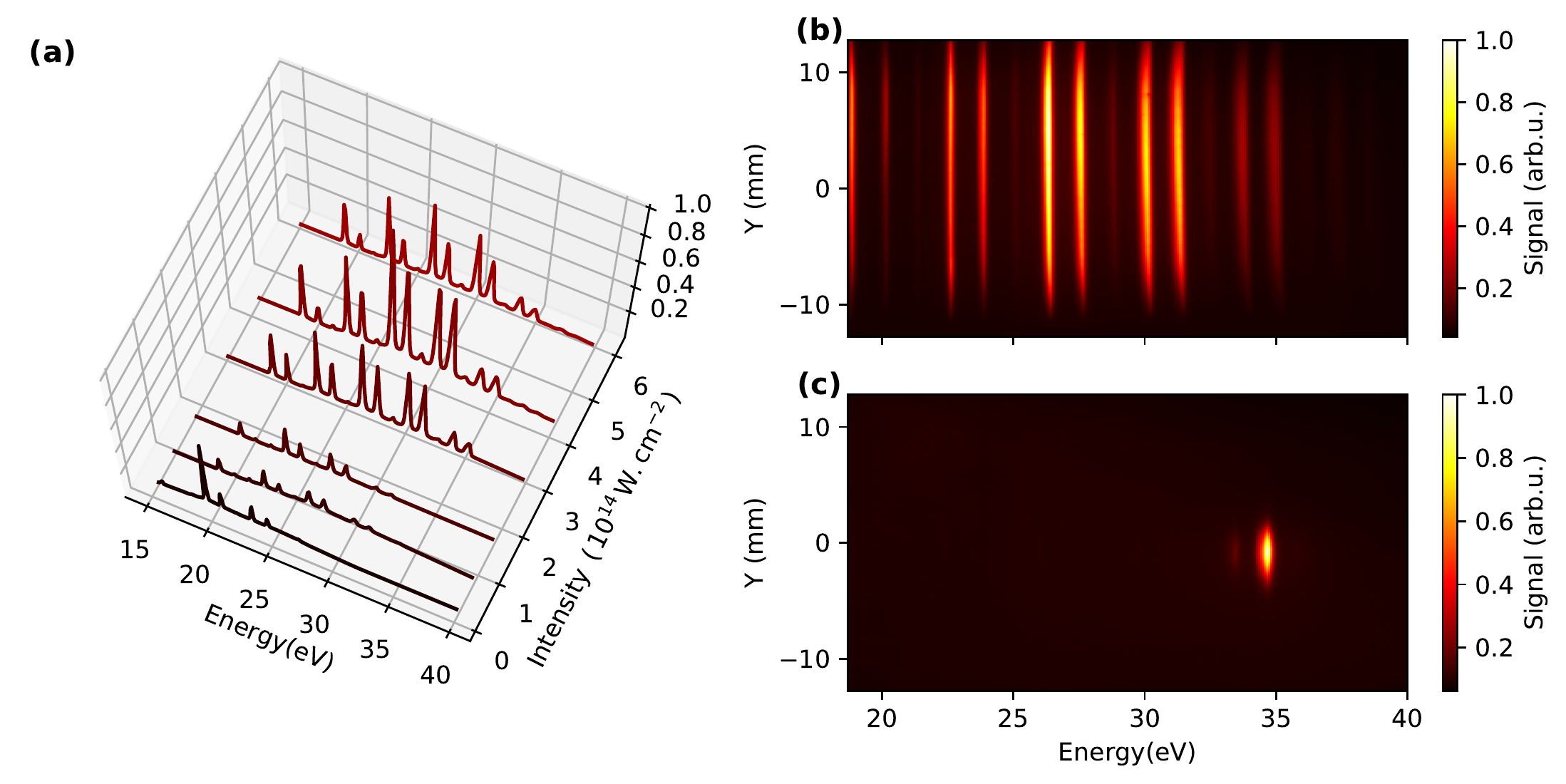}
\caption{Bichromatic bicircular HHG. (a) Intensity dependency of the HHG spectrum. Spatially resolved bichromatic bicircular HHG spectrum before (b) and after (c) multilayer mirrors at 35 eV, showing both the spectral selection and refocusing. }
\label{figmazeltov}
\end{center}
\end{figure}

To produce the bicircular high-order harmonics, we used an in-line setup called Mazel-Tov, developed by Kfir \textit{et al.} on Ti:Sa systems \cite{kfir16}. The 1030 nm beam with 50 W average power (3.7 mm waist FWHM) goes through a 1-mm BBO crystal to generate 15 W of  second harmonic, which is orthogonally polarized with respect to the remaining 35 W of 1030 nm. The temporal overlap of the two beams is set by propagation in a 1-mm calcite crystal, and finely adjusted through propagation into a pair of thin fused SiO$_2$ wedges (about 2 mm length, finely tunable). A superachromatic quarter waveplate (300-1100 nm) converts the two orthogonal linear polarizations into two counter-rotating circular polarizations. The bichromatic bicircular beam is then focused in the HHG chamber by two dual band dielectric mirrors in order to avoid chromatic aberration. Due to the tight focusing condition and the need to work at $0^{\circ}$ incidence to preserve polarization states, the focusing optics are set under vacuum, close to the gas jet.
The intensity at focus can be tuned either by  decreasing the focal length of the focusing mirror (typically 150 mm), or by enlarging the beam before the quarter waveplate with a telescope made of dual band mirror (typical magnification $\times 2$).

We present the spectrum obtained at different intensities at 515 nm in Fig. \ref{figmazeltov} (a). The first spectrum was obtained at low intensity (about $2 \times 10^{13}$ W/cm$^2$), resulting in a low cutoff (24 eV), way below the targeted 35 eV. We thus increased the intensity by changing both the telescope and the focusing mirror in order to increase $U_p$ and thus the cutoff ($E_{max}=I_p+3.17 U_p$ \cite{corkum93}). We observe a clear saturation of the cutoff around 35 eV  \linebreak  when the laser intensity reaches $3 \times 10^{14}$ W/cm$^2$. 
We interprete this as the result of the saturation of ionization by the 515 nm \cite{paul06,kazamias11}. This limitation is intrinsic to the relatively long duration of our laser pulses (135 fs). Nevertheless, the spatially resolved spectrum shown in Fig. \ref{figmazeltov} (b) has a clean spatial profile, with significant signal at 35 eV. \linebreak  In Fig. \ref{figmazeltov} (c), we present the spectrum obtained after the two multilayer mirrors. Harmonic 29 is clearly dominant with a estimated flux of $10^{10}$ photons/s. This scheme thus opens the route to photoelectron circular dichroism experiments at this photon energy.

\section{Coincidence electron-ion imaging}
Coincidence electron-ion detection is a very powerful spectroscopic tool, with many applications in a wide variety of fields, ranging from channel-resolved strong field ionization \cite{boguslavskiy12}, molecular orbital imaging \cite{sann16}, state-resolved photoelectron circular dichroism in chiral molecules \cite{fehre19}, to the measurement of orientation-dependent attosecond Wigner delays in molecules \cite{vos18}. 
The double electron-ion imaging offered by COLTRIMS detectors enables for instance distinguishing the photoelectron spectra from monomers and clusters \cite{powis14}, and to record molecular-frame photoelectron angular distributions \cite{lucchese12,tia17}. In addition to this coincidence capacity, the use of delay-line anodes in the COLTRIMS also enables the direct measurement of 3D photoelectron angular distributions, without any symmetry asumption as required in velocity map imaging. We recently used this property, combined with the benefit of the high repetition rate of our YDFA laser, to record 3D photoelectron angular distributions of chiral molecules ionized by an elliptically polarized laser field \cite{comby18}, or to investigate chiral photoionization by tailored laser fields with sub-cycle chirality variations \cite{rozen19}. While such laser-based studies are of undeniable interest, single-photon ionization by XUV radiation remains the spectroscopic method of choice for many investigations. In this section we thus present the coupling of our cascaded HHG XUV source with the coincidence imaging electron-ion spectrometer. 

As an illustration of the potential of the beamline, we studied the XUV photoionization of xenon atoms and clusters. High-order harmonics were produced using 7.5 W of 515 nm radiation at 166 kHz, and argon as the generating gas, leading to a flux on target after the SiC mirrors of $\sim 10^{10}$ photons/s, as presented in section 4. This flux was sufficient to reach a 15 kHz count rate in the COLTRIMS, leaving a significant margin of improvement in the case where processes with lower cross-sections would be investigated. No elements were required here to filter out the $2 \omega_{L}$ field as the 3 m focal length of the refocusing mirror prevents the appearance of multiphoton processes.

\begin{figure}[h]
\begin{center}
\includegraphics[width=\textwidth]{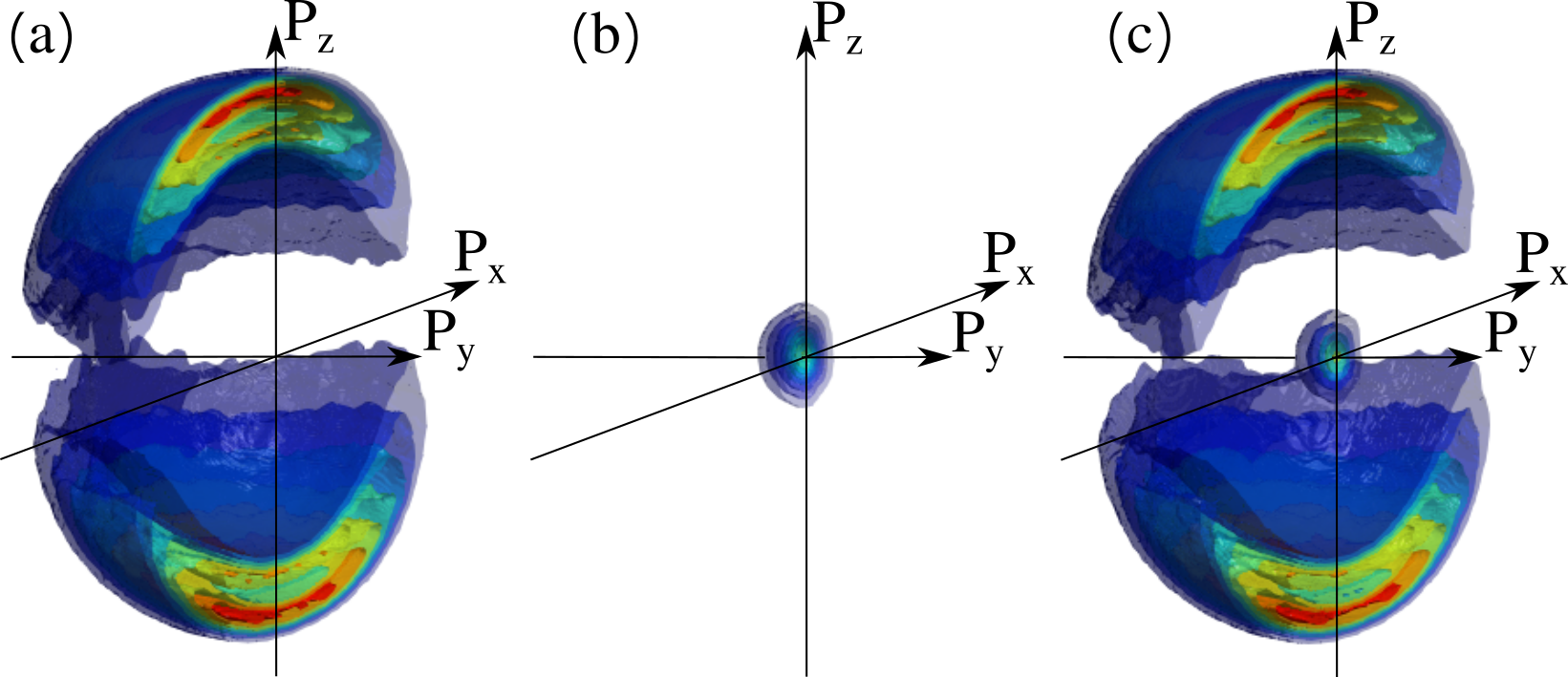}
\caption{Halves of the 3D photoelectron angular distribution recorded by ionizing xenon with high harmonics of 515 nm pulses. Distributions associated with Xe monomers (a), Xe dimers (b) and without coincidence filter (c). The harmonics are linearly polarized along z and propagate along x.}
\label{3D_PES}
\end{center}
\end{figure}

Part of the events acquired are not usable in the final photoelectron angular distribution for several reasons (noticeably the ion detector efficiency, the events coming from the gas background in the interaction chamber and the events that come from a too wide spatial extension of the beam along its propagation axis due to the lose focusing). Moreover, at high repetition rate, the time between two consecutive laser shots becomes shorter than the ion time-of-flight (about 12 $\mu$s with Xe$^+$ in our configuration). This can increase the probability to record two events at a time without having the possibility to unambiguously assign the corresponding electrons and ions. This can strengthen the limit on the ionization per shot probability when using high repetition rates. 150M counts were recorded in total in about 3 hours. Only 1.5M of them have been used here as a damage on the ion MCPs limited the detection efficiency. We estimate that replacing the detector would provide us with 30 times more usable signal. 

Fig. \ref{3D_PES} shows the photoelectron angular distribution experimentally measured, respectively in coincidence with the Xe monomers (a), the Xe dimers (b) and without coincidence filter (c). In the clustering conditions used (the jet was produced by backing the nozzle with 1 bar of xenon at room temperature), the ratio between monomer and dimer counts was about 100:1. Photoelectrons associated with Xe monomers represent 97.8$\%$ of the signal, and appear with two peaks of kinetic energy centered at 3.3 eV and 4.6 eV. This corresponds to the ionization to the $5s^{2}5p^{5}$ $^{2}P^{0}_{\frac{1}{2}}$ and $5s^{2}5p^{5}$ $^{2}P^{0}_{\frac{3}{2}}$ states (of respective ionization thresholds of 13.44 eV and 12.13 eV) by the 7$^{th}$ harmonic of the 515 nm field (centered at 16.85 eV). The electrons produced by Xe dimers (2.2$\%$ of the signal)  show a kinetic energy distribution centered at 0.15 eV, and are produced by absorption of the 5$^{th}$ harmonic centered at 12.04 eV, revealing the well-known $\sim -0.24$ eV  \linebreak  shift in the Ip of the dimer compared to the monomer \cite{lu_ground_1995}.

This example typically illustrates the capacity of the photoelectron-photoion coincidence imaging detection to disentangle the different contributions in a photoelectron distribution, revealing the signal from a minority species in a mixture.

\section{Conclusion}
The combination of robust, turn-key laser technology with simple non-linear optics and high-order harmonic generation results in an efficient and versatile source of XUV radiation. The cascaded HHG process enables very high conversion efficiencies, producing up to 1.8 mW of radiation. Various spectral selection configurations can be implemented for different applications. In particular, the generation of circularly polarized radiation at 35 eV, associated with the coincidence electron-ion imaging of the COLTRIMS, opens the way to a broad range of investigations in time-resolved photoelectron circular dichroism \cite{comby16}.

%\section*{Acknowledgments}
\ack
We thank R. Bouillaud, F. Blais, N. Fedorov, A. Filippov and L. Merzeau for technical assistance, D. Azoury, B. Bruner and J. Gaudin for experimental support, and S. Patchkovskii and B. Pons for fruitful discussion. This project has received funding from the European Research Council (ERC) under the European Union’s Horizon 2020 research and innovation programme no. 682978 - EXCITERS. We acknowledge financial support of the the French National Research Agency through ANR-14-CE32-0014 MISFITS.

\section*{References}
\bibliographystyle{unsrt}
\bibliography{2020-HHG-20200309.bib}

\begin{thebibliography}{10}

\bibitem{mcpherson87}
A.~McPherson, G.~Gibson, H.~Jara, U.~Johann, T.~S. Luk, I.~McIntyre, K.~Boyer,
  and C.~K. Rhodes.
\newblock Studies of multiphoton production of vacuum-ultraviolet radiation in
  the rare gases.
\newblock {\em J. Opt. Soc. Am. B}, 4:595, 1987.

\bibitem{ferray88}
M~Ferray, A~L'Huillier, X~F Li, L~A Lompre, G~Mainfray, and C~Manus.
\newblock Multiple-harmonic conversion of 1064 nm radiation in rare gases.
\newblock {\em Journal of Physics B: Atomic, Molecular and Optical Physics},
  21(3):L31--L35, February 1988.

\bibitem{lhuillier03}
A.~L'Huillier, A.~Johansson, J.~Norin, J.~Mauritsson, and C.-G. Wahlstr\"om.
\newblock Applications of high-order harmonics.
\newblock {\em Eur. Phys. J. D}, 26:91, 2003.

\bibitem{krausz09}
Ferenc Krausz and Misha Ivanov.
\newblock Attosecond physics.
\newblock {\em Reviews of Modern Physics}, 81(1):163--234, February 2009.

\bibitem{lepine14}
Franck L\'epine, Misha~Y. Ivanov, and Marc J.~J. Vrakking.
\newblock Attosecond molecular dynamics: fact or fiction?
\newblock {\em Nature Photonics}, 8(3):195--204, March 2014.

\bibitem{nisoli17}
Mauro Nisoli, Piero Decleva, Francesca Calegari, Alicia Palacios, and Fernando
  Martín.
\newblock Attosecond {Electron} {Dynamics} in {Molecules}.
\newblock {\em Chemical Reviews}, 117(16):10760--10825, August 2017.

\bibitem{hergott02}
J.-F. Hergott, M.~Kovacev, H.~Merdji, C.~Hubert, Y.~Mairesse, E.~Jean,
  P.~Breger, P.~Agostini, B.~Carr\'e, and P.~Sali\`eres.
\newblock Extreme-ultraviolet high-order harmonic pulses in the microjoule
  range.
\newblock {\em Physical Review A}, 66(2):021801, 2002.

\bibitem{takahashi02}
Eiji Takahashi, Yasuo Nabekawa, and Katsumi Midorikawa.
\newblock Generation of 10-µ{J} coherent extreme-ultraviolet light by use of
  high-order harmonics.
\newblock {\em Optics Letters}, 27(21):1920--1922, November 2002.

\bibitem{tzallas03}
P.~Tzallas, D.~Charalambidis, N.~A. Papadogiannis, K.~Witte, and G.~D.
  Tsakiris.
\newblock Direct observation of attosecond light bunching.
\newblock {\em Nature}, 426(6964):267--271, 2003.

\bibitem{takahashi13}
Eiji~J. Takahashi, Pengfei Lan, Oliver~D. Mücke, Yasuo Nabekawa, and Katsumi
  Midorikawa.
\newblock Attosecond nonlinear optics using gigawatt-scale isolated attosecond
  pulses.
\newblock {\em Nature Communications}, 4(1):1--9, October 2013.

\bibitem{senfftleben19}
Bj\"orn Senfftleben, Martin Kretschmar, Andreas Hoffmann, Mario Sauppe,
  Johannes Tümmler, Ingo Will, Tamás Nagy, Marc J.~J. Vrakking, Daniela Rupp,
  and Bernd Schütte.
\newblock Highly nonlinear ionization of atoms induced by intense high-harmonic
  pulses.
\newblock {\em arXiv:1911.01375 [physics]}, November 2019.
\newblock arXiv: 1911.01375.

\bibitem{nisoli96}
M.~Nisoli, S.~De Silvestri, and O.~Svelto.
\newblock Generation of high energy 10 fs pulses by a new pulse compression
  technique.
\newblock {\em Appl. Phys. Lett.}, 68:2793, 1996.

\bibitem{xu_route_1996}
L.~Xu, Ch~Spielmann, A.~Poppe, T.~Brabec, F.~Krausz, and T.~W. H\"ansch.
\newblock Route to phase control of ultrashort light pulses.
\newblock {\em Optics Letters}, 21(24):2008--2010, December 1996.

\bibitem{hentschel01}
M.~Hentschel, R.~Kienberger, Ch. Spielmann, G.~A. Reider, N.~Milosevic,
  T.~Brabec, P.~Corkum, U.~Heinzmann, M.~Drescher, and F.~Krausz.
\newblock Attosecond metrology.
\newblock {\em Nature}, 414:509, 2001.

\bibitem{chini14}
Michael Chini, Kun Zhao, and Zenghu Chang.
\newblock The generation, characterization and applications of broadband
  isolated attosecond pulses.
\newblock {\em Nature Photonics}, 8(3):178--186, March 2014.

\bibitem{chen10}
M.-C. Chen, P.~Arpin, T.~Popmintchev, M.~Gerrity, B.~Zhang, M.~Seaberg,
  D.~Popmintchev, M.~M. Murnane, and H.~C. Kapteyn.
\newblock Bright, {Coherent}, {Ultrafast} {Soft} {X}-{Ray} {Harmonics}
  {Spanning} the {Water} {Window} from a {Tabletop} {Light} {Source}.
\newblock {\em Physical Review Letters}, 105(17):173901, October 2010.

\bibitem{ren18}
Xiaoming Ren, Jie Li, Yanchun Yin, Kun Zhao, Andrew Chew, Yang Wang, Shuyuan
  Hu, Yan Cheng, {Eric Cunningham}, Yi~Wu, Michael Chini, and Zenghu Chang.
\newblock Attosecond light sources in the water window.
\newblock {\em Journal of Optics}, 20(2):023001, 2018.

\bibitem{ullrich03}
J.~Ullrich, R.~Moshammer, A.~Dorn, R.~D\"orner, L.P.H. Schmidt, and
  H.~Schmidt-B\"ocking.
\newblock Recoil-ion and electron momentum spectroscopy: reaction-microscopes.
\newblock {\em Reports on Progress in Physics}, 66:1463, 2003.

\bibitem{schonhense18}
B.~Sch\"onhense, K.~Medjanik, O.~Fedchenko, S.~Chernov, M.~Ellguth,
  D.~Vasilyev, A.~Oelsner, J.~Viefhaus, D.~Kutnyakhov, W.~Wurth, H.~J. Elmers,
  and G.~Sch\"onhense.
\newblock Multidimensional photoemission spectroscopy—the space-charge limit.
\newblock {\em New Journal of Physics}, 20(3):033004, March 2018.
\newblock Publisher: IOP Publishing.

\bibitem{corder18}
Christopher Corder, Peng Zhao, Jin Bakalis, Xinlong Li, Matthew~D. Kershis,
  Amanda~R. Muraca, Michael~G. White, and Thomas~K. Allison.
\newblock Ultrafast extreme ultraviolet photoemission without space charge.
\newblock {\em Structural Dynamics}, 5(5):054301, September 2018.
\newblock Publisher: American Institute of Physics.

\bibitem{saule19}
T.~Saule, S.~Heinrich, J.~Sch\"otz, N.~Lilienfein, M.~H\"ogner, O.~deVries,
  M.~Pl\"otner, J.~Weitenberg, D.~Esser, J.~Schulte, P.~Russbueldt, J.~Limpert,
  M.~F. Kling, U.~Kleineberg, and I.~Pupeza.
\newblock High-flux ultrafast extreme-ultraviolet photoemission spectroscopy at
  18.4 {MHz} pulse repetition rate.
\newblock {\em Nature Communications}, 10(1):1--10, January 2019.
\newblock Number: 1 Publisher: Nature Publishing Group.

\bibitem{russbueldt10}
P.~Russbueldt, T.~Mans, J.~Weitenberg, H.~D. Hoffmann, and R.~Poprawe.
\newblock Compact diode-pumped 1.1 kw yb:yag innoslab femtosecond amplifier.
\newblock {\em Opt. Lett.}, 35(24):4169--4171, Dec 2010.

\bibitem{eidam10}
Tino Eidam, Stefan Hanf, Enrico Seise, Thomas~V. Andersen, Thomas Gabler,
  Christian Wirth, Thomas Schreiber, Jens Limpert, and Andreas T\"{u}nnermann.
\newblock Femtosecond fiber cpa system emitting 830 w average output power.
\newblock {\em Opt. Lett.}, 35(2):94--96, Jan 2010.

\bibitem{ruehl10}
Axel Ruehl, Andrius Marcinkevicius, Martin~E. Fermann, and Ingmar Hartl.
\newblock 80 w, 120 fs yb-fiber frequency comb.
\newblock {\em Opt. Lett.}, 35(18):3015--3017, Sep 2010.

\bibitem{lavenu17}
L.~Lavenu, M.~Natile, F.~Guichard, Y.~Zaouter, M.~Hanna, E.~Mottay, and
  P.~Georges.
\newblock High-energy few-cycle {Yb}-doped fiber amplifier source based on a
  single nonlinear compression stage.
\newblock {\em Optics Express}, 25(7):7530, April 2017.

\bibitem{boullet09}
Johan Boullet, Yoann Zaouter, Jens Limpert, St\'ephane Petit, Yann Mairesse,
  Baptiste Fabre, Julien Higuet, Eric M\'evel, Eric Constant, and Eric Cormier.
\newblock High-order harmonic generation at a megahertz-level repetition rate
  directly driven by an ytterbium-doped-fiber chirped-pulse amplification
  system.
\newblock {\em Opt. Lett.}, 34:1489, 2009.

\bibitem{rothhardt14-2}
Jan Rothhardt, Manuel Krebs, Steffen H\"adrich, Stefan Demmler, Jens Limpert,
  and Andreas Tünnermann.
\newblock Absorption-limited and phase-matched high harmonic generation in the
  tight focusing regime.
\newblock {\em New Journal of Physics}, 16(3):033022, 2014.

\bibitem{heyl16}
C.~M. Heyl, H.~Coudert-Alteirac, M.~Miranda, M.~Louisy, K.~Kovacs, V.~Tosa,
  E.~Balogh, K.~Varjú, A.~L’Huillier, A.~Couairon, and C.~L. Arnold.
\newblock Scale-invariant nonlinear optics in gases.
\newblock {\em Optica}, 3(1):75, January 2016.

\bibitem{wang15}
He~Wang, Yiming Xu, Stefan Ulonska, Joseph~S. Robinson, Predrag Ranitovic, and
  Robert~A. Kaindl.
\newblock Bright high-repetition-rate source of narrowband extreme-ultraviolet
  harmonics beyond 22 {eV}.
\newblock {\em Nature Communications}, 6, June 2015.

\bibitem{klas16}
R.~Klas, S.~Demmler, M.~Tschernajew, S.~H\"adrich, Y.~Shamir, A.~Tünnermann,
  J.~Rothhardt, and J.~Limpert.
\newblock Table-top milliwatt-class extreme ultraviolet high harmonic light
  source.
\newblock {\em Optica}, 3(11):1167, November 2016.

\bibitem{porat18-1}
Gil Porat, Christoph~M. Heyl, Stephen~B. Schoun, Craig Benko, Nadine D\"orre,
  Kristan~L. Corwin, and Jun Ye.
\newblock Phase-matched extreme-ultraviolet frequency-comb generation.
\newblock {\em Nature Photonics}, 12(7):387, July 2018.

\bibitem{ritchie76}
Burke Ritchie.
\newblock Theory of the angular distribution of photoelectrons ejected from
  optically active molecules and molecular negative ions.
\newblock {\em Phys. Rev. A}, 13(4):1411--1415, April 1976.

\bibitem{bowering01}
N.~B\"owering, T.~Lischke, B.~Schmidtke, N.~Müller, T.~Khalil, and
  U.~Heinzmann.
\newblock Asymmetry in {Photoelectron} {Emission} from {Chiral} {Molecules}
  {Induced} by {Circularly} {Polarized} {Light}.
\newblock {\em Physical Review Letters}, 86(7):1187--1190, February 2001.

\bibitem{nahon15}
Laurent Nahon, Gustavo~A. Garcia, and Ivan Powis.
\newblock Valence shell one-photon photoelectron circular dichroism in chiral
  systems.
\newblock {\em Journal of Electron Spectroscopy and Related Phenomena}, 204,
  Part B:322--334, October 2015.

\bibitem{comby19}
A.~Comby, D.~Descamps, S.~Beauvarlet, A.~Gonzalez, F.~Guichard, S.~Petit,
  Y.~Zaouter, and Y.~Mairesse.
\newblock Cascaded harmonic generation from a fiber laser: a milliwatt {XUV}
  source.
\newblock {\em Optics Express}, 27(15):20383--20396, July 2019.

\bibitem{comby18}
A.~Comby, S.~Beaulieu, E.~Constant, D.~Descamps, S.~Petit, and Y.~Mairesse.
\newblock Absolute gas density profiling in high-order harmonic generation.
\newblock {\em Optics Express}, 26(5):6001--6009, March 2018.

\bibitem{horke17}
Daniel~A. Horke, Nils Roth, Lena Worbs, and Jochen Küpper.
\newblock Characterizing gas flow from aerosol particle injectors.
\newblock {\em Journal of Applied Physics}, 121(12):123106, 2017.

\bibitem{balcou92}
P.~Balcou, C.~Cornaggia, A.~S.~L. Gomes, L.~A. Lompre, and
  A.~L{\textbackslash}textquotesingleHuillier.
\newblock Optimizing high-order harmonic generation in strong fields.
\newblock {\em Journal of Physics B: Atomic, Molecular and Optical Physics},
  25(21):4467--4485, November 1992.

\bibitem{colosimo08}
P.~Colosimo, G.~Doumy, C.~I. Blaga, J.~Wheeler, C.~Hauri, F.~Catoire, J.~Tate,
  R.~Chirla, A.~M. March, G.~G. Paulus, H.~G. Muller, P.~Agostini, and L.~F.
  DiMauro.
\newblock Scaling strong-field interactions towards the classical limit.
\newblock {\em Nature Physics}, 4(5):386--389, March 2008.

\bibitem{shiner09}
A.~D. Shiner, C.~Trallero-Herrero, N.~Kajumba, H.-C. Bandulet, D.~Comtois,
  F.~L\'egar\'e, M.~Gigu\`ere, J-C. Kieffer, P.~B. Corkum, and D.~M.
  Villeneuve.
\newblock Wavelength {Scaling} of {High} {Harmonic} {Generation} {Efficiency}.
\newblock {\em Physical Review Letters}, 103(7):073902, 2009.

\bibitem{lai13}
Chien-Jen Lai, Giovanni Cirmi, Kyung-Han Hong, Jeffrey Moses, Shu-Wei Huang,
  Eduardo Granados, Phillip Keathley, Siddharth Bhardwaj, and Franz~X.
  K\"artner.
\newblock Wavelength {Scaling} of {High} {Harmonic} {Generation} {Close} to the
  {Multiphoton} {Ionization} {Regime}.
\newblock {\em Physical Review Letters}, 111(7):073901, August 2013.

\bibitem{marceau17}
Claude Marceau, Varun Makhija, Dominique Platzer, A.~Yu. Naumov, P. B.
  Corkum, Albert Stolow, D. M. Villeneuve, and Paul Hockett.
\newblock Molecular {Frame} {Reconstruction} {Using} {Time}-{Domain}
  {Photoionization} {Interferometry}.
\newblock {\em Physical Review Letters}, 119(8):083401, August 2017.

\bibitem{popmintchev15}
Dimitar Popmintchev, Carlos Hernández-García, Franklin Dollar, Christopher
  Mancuso, Jose~A. P\'erez-Hernández, Ming-Chang Chen, Amelia Hankla, Xiaohui
  Gao, Bonggu Shim, Alexander~L. Gaeta, Maryam Tarazkar, Dmitri~A. Romanov,
  Robert~J. Levis, Jim~A. Gaffney, Mark Foord, Stephen~B. Libby, Agnieszka
  Jaron-Becker, Andreas Becker, Luis Plaja, Margaret~M. Murnane, Henry~C.
  Kapteyn, and Tenio Popmintchev.
\newblock Ultraviolet surprise: {Efficient} soft x-ray high-harmonic generation
  in multiply ionized plasmas.
\newblock {\em Science}, 350(6265):1225--1231, December 2015.

\bibitem{klas18}
R.~Klas, A.~Kirsche, M.~Tschernajew, J.~Rothhardt, and J.~Limpert.
\newblock Annular beam driven high harmonic generation for high flux coherent
  {XUV} and soft {X}-ray radiation.
\newblock {\em Optics Express}, 26(15):19318--19327, July 2018.

\bibitem{heyl17}
C.~M. Heyl, C.~L. Arnold, A.~Couairon, and A.~L’Huillier.
\newblock Introduction to macroscopic power scaling principles for high-order
  harmonic generation.
\newblock {\em Journal of Physics B: Atomic, Molecular and Optical Physics},
  50(1):013001, 2017.

\bibitem{porat18}
G.~Porat, G.~Alon, S.~Rozen, O.~Pedatzur, M.~Krüger, D.~Azoury, A.~Natan,
  G.~Orenstein, B.~D. Bruner, M.~J.~J. Vrakking, and N.~Dudovich.
\newblock Attosecond time-resolved photoelectron holography.
\newblock {\em Nature Communications}, 9(1):2805, July 2018.

\bibitem{manschwetus16}
B.~Manschwetus, L.~Rading, F.~Campi, S.~Maclot, H.~Coudert-Alteirac, J.~Lahl,
  H.~Wikmark, P.~Rudawski, C.~M. Heyl, B.~Farkas, T.~Mohamed, A.~L'Huillier,
  and P.~Johnsson.
\newblock Two-photon double ionization of neon using an intense attosecond
  pulse train.
\newblock {\em Physical Review A}, 93(6):061402, June 2016.

\bibitem{gauthier10}
D.~Gauthier, M.~Guizar-Sicairos, X.~Ge, W.~Boutu, B.~Carr\'e, J.~R. Fienup, and
  H.~Merdji.
\newblock Single-shot {Femtosecond} {X}-{Ray} {Holography} {Using} {Extended}
  {References}.
\newblock {\em Physical Review Letters}, 105(9):093901, August 2010.

\bibitem{constant99}
E.~Constant, D.~Garzella, P.~Breger, E.~Mevel, C.~Dorrer, C.~Le~Blanc,
  F.~Salin, and P.~Agostini.
\newblock Optimizing high harmonic generation in absorbing gases: {Model} and
  experiment.
\newblock {\em Physical review letters}, 82(8):1668--1671, 1999.

\bibitem{ding14}
Chengyuan Ding, Wei Xiong, Tingting Fan, Daniel~D. Hickstein, Tenio
  Popmintchev, Xiaoshi Zhang, Mike Walls, Margaret~M. Murnane, and Henry~C.
  Kapteyn.
\newblock High flux coherent super-continuum soft {X}-ray source driven by a
  single-stage, 10mj, {Ti}:sapphire amplifier-pumped {OPA}.
\newblock {\em Optics Express}, 22(5):6194--6202, March 2014.

\bibitem{cabasse12}
Am\'elie Cabasse, Guillaume Machinet, Antoine Dubrouil, Eric Cormier, and Eric
  Constant.
\newblock Optimization and phase matching of fiber-laser-driven high-order
  harmonic generation at high repetition rate.
\newblock {\em Optics Letters}, 37(22):4618, November 2012.

\bibitem{rothhardt14}
Jan Rothhardt, Steffen H\"adrich, Stefan Demmler, Manuel Krebs, Stephan
  Fritzsche, Jens Limpert, and Andreas Tünnermann.
\newblock Enhancing the {Macroscopic} {Yield} of {Narrow}-{Band} {High}-{Order}
  {Harmonic} {Generation} by {Fano} {Resonances}.
\newblock {\em Physical Review Letters}, 112(23):233002, June 2014.

\bibitem{hadrich14}
Steffen H\"adrich, Arno Klenke, Jan Rothhardt, Manuel Krebs, Armin Hoffmann,
  Oleg Pronin, Vladimir Pervak, Jens Limpert, and Andreas Tünnermann.
\newblock High photon flux table-top coherent extreme-ultraviolet source.
\newblock {\em Nature Photonics}, 8(10):779--783, October 2014.

\bibitem{lee11}
Jane Lee, David~R. Carlson, and R.~Jason Jones.
\newblock Optimizing intracavity high harmonic generation for {XUV} fs
  frequency combs.
\newblock {\em Optics Express}, 19(23):23315--23326, November 2011.

\bibitem{cingoz12}
Arman Cing\"oz, Dylan~C. Yost, Thomas~K. Allison, Axel Ruehl, Martin~E.
  Fermann, Ingmar Hartl, and Jun Ye.
\newblock Direct frequency comb spectroscopy in the extreme ultraviolet.
\newblock {\em Nature}, 482(7383):68--71, February 2012.

\bibitem{pupeza13}
I.~Pupeza, S.~Holzberger, T.~Eidam, H.~Carstens, D.~Esser, J.~Weitenberg,
  P.~Rußbüldt, J.~Rauschenberger, J.~Limpert, Th~Udem, A.~Tünnermann, T.~W.
  H\"ansch, A.~Apolonski, F.~Krausz, and E.~Fill.
\newblock Compact high-repetition-rate source of coherent 100 {eV} radiation.
\newblock {\em Nature Photonics}, 7(8):608--612, August 2013.

\bibitem{tigrine16}
S~Tigrine, N~Carrasco, L~Vettier, and G~Cernogora.
\newblock A microwave plasma source for {VUV} atmospheric photochemistry.
\newblock {\em Journal of Physics D: Applied Physics}, 49(39):395202, sep 2016.

\bibitem{nahon13}
L~Nahon, N~de Oliveira, G~A Garcia, J-F Gil, D~Joyeux, B~Lagarde, and F~Polack.
\newblock {DESIRS} : a state-of-the-art {VUV} beamline featuring high
  resolution and variable polarization for spectroscopy and dichroism at
  {SOLEIL}.
\newblock {\em Journal of Physics: Conference Series}, 425(12):122004, March
  2013.

\bibitem{bourgalais19}
J\'er\'emy Bourgalais, Nathalie Carrasco, Ludovic Vettier, Thomas Gautier,
  Val\'erie Blanchet, St\'ephane Petit, Dominique Descamps~Nikita Fedorov,
  Romain Delos, and J\'erôme Gaudin.
\newblock An {EUV} {Non}-{Linear} {Optics} {Based} {Approach} to {Study} the
  {Photochemical} {Processes} of {Titan}{\textbackslash}'s {Atmosphere}.
\newblock {\em arXiv:1910.00362 [astro-ph]}, October 2019.
\newblock arXiv: 1910.00362.

\bibitem{barreau18}
Lou Barreau, K\'evin Veyrinas, Vincent Gruson, S\'ebastien~J. Weber, Thierry
  Auguste, Jean-François Hergott, Fabien Lepetit, Bertrand Carr\'e,
  Jean-Christophe Houver, Danielle Dowek, and Pascal Sali\`eres.
\newblock Evidence of depolarization and ellipticity of high harmonics driven
  by ultrashort bichromatic circularly polarized fields.
\newblock {\em Nature Communications}, 9(1):4727, November 2018.

\bibitem{milosevic00}
D.B. Milosevic.
\newblock {\em J. Phys. B}, 33:2479, 2000.

\bibitem{kfir16}
Ofer Kfir, Eliyahu Bordo, Gil Ilan~Haham, Oren Lahav, Avner Fleischer, and Oren
  Cohen.
\newblock In-line production of a bi-circular field for generation of helically
  polarized high-order harmonics.
\newblock {\em Applied Physics Letters}, 108(21):211106, May 2016.

\bibitem{vanderlaan14}
Gerrit van~der Laan and Adriana~I. Figueroa.
\newblock X-ray magnetic circular dichroism—{A} versatile tool to study
  magnetism.
\newblock {\em Coordination Chemistry Reviews}, 277-278:95--129, October 2014.

\bibitem{vodungbo11}
Boris Vodungbo, Anna Barszczak~Sardinha, Julien Gautier, Guillaume Lambert,
  Constance Valentin, Magali Lozano, Gr\'egory Iaquaniello, Franck Delmotte,
  St\'ephane Sebban, Jan Lüning, and Philippe Zeitoun.
\newblock Polarization control of high order harmonics in the {EUV} photon
  energy range.
\newblock {\em Optics Express}, 19(5):4346--4356, February 2011.

\bibitem{willems15}
F.~Willems, C.~T.~L. Smeenk, N.~Zhavoronkov, O.~Kornilov, I.~Radu,
  M.~Schmidbauer, M.~Hanke, C.~von Korff~Schmising, M.~J.~J. Vrakking, and
  S.~Eisebitt.
\newblock Probing ultrafast spin dynamics with high-harmonic magnetic circular
  dichroism spectroscopy.
\newblock {\em Physical Review B}, 92(22):220405, December 2015.

\bibitem{siegrist19}
Florian Siegrist, Julia~A. Gessner, Marcus Ossiander, Christian Denker, Yi-Ping
  Chang, Malte~C. Schr\"oder, Alexander Guggenmos, Yang Cui, Jakob Walowski,
  Ulrike Martens, J.~K. Dewhurst, Ulf Kleineberg, Markus Münzenberg, Sangeeta
  Sharma, and Martin Schultze.
\newblock Light-wave dynamic control of magnetism.
\newblock {\em Nature}, 571(7764):240--244, July 2019.
\newblock Number: 7764 Publisher: Nature Publishing Group.

\bibitem{azoury19}
Doron Azoury, Omer Kneller, Michael Krüger, Barry~D. Bruner, Oren Cohen, Yann
  Mairesse, and Nirit Dudovich.
\newblock Interferometric attosecond lock-in measurement of extreme-ultraviolet
  circular dichroism.
\newblock {\em Nature Photonics}, 13(3):198, March 2019.

\bibitem{ferre15-1}
A.~Ferr\'e, C.~Handschin, M.~Dumergue, F.~Burgy, A.~Comby, D.~Descamps,
  B.~Fabre, G.~A. Garcia, R.~G\'eneaux, L.~Merceron, E.~M\'evel, L.~Nahon,
  S.~Petit, B.~Pons, D.~Staedter, S.~Weber, T.~Ruchon, V.~Blanchet, and
  Y.~Mairesse.
\newblock A table-top ultrashort light source in the extreme ultraviolet for
  circular dichroism experiments.
\newblock {\em Nature Photonics}, 9(2):93--98, February 2015.

\bibitem{muth-bohm00}
J.~Muth-B\"ohm, A.~Becker, and F.~H.~M. Faisal.
\newblock Suppressed {Molecular} {Ionization} for a {Class} of {Diatomics} in
  {Intense} {Femtosecond} {Laser} {Fields}.
\newblock {\em Physical Review Letters}, 85(11):2280--2283, September 2000.

\bibitem{tong02}
X.~M. Tong, Z.~X. Zhao, and C.~D. Lin.
\newblock Theory of molecular tunneling ionization.
\newblock {\em Physical Review A}, 66(3), September 2002.

\bibitem{ferre15}
A.~Ferr\'e, A.~E. Boguslavskiy, M.~Dagan, V.~Blanchet, B.~D. Bruner, F.~Burgy,
  A.~Camper, D.~Descamps, B.~Fabre, N.~Fedorov, J.~Gaudin, G.~Geoffroy,
  J.~Mikosch, S.~Patchkovskii, S.~Petit, T.~Ruchon, H.~Soifer, D.~Staedter,
  I.~Wilkinson, A.~Stolow, N.~Dudovich, and Y.~Mairesse.
\newblock Multi-channel electronic and vibrational dynamics in polyatomic
  resonant high-order harmonic generation.
\newblock {\em Nature Communications}, 6, January 2015.

\bibitem{antoine97-2}
Philippe Antoine, Bertrand Carr\'e, Anne L'Huillier, and Maciej Lewenstein.
\newblock Polarization of high-order harmonics.
\newblock {\em Physical Review A}, 55(2):1314--1324, February 1997.

\bibitem{veyrinas16}
K.~Veyrinas, V.~Gruson, S.~J. Weber, L.~Barreau, T.~Ruchon, J.-F. Hergott,
  J.-C. Houver, R.~R. Lucchese, P.~Sali\`eres, and D.~Dowek.
\newblock Molecular frame photoemission by a comb of elliptical high-order
  harmonics: a sensitive probe of both photodynamics and harmonic complete
  polarization state.
\newblock {\em Faraday Discussions}, 194(0):161--183, December 2016.

\bibitem{nahon04}
Laurent Nahon and Christian Alcaraz.
\newblock {SU}5: a calibrated variable-polarization synchrotron radiation beam
  line in the vacuum-ultraviolet range.
\newblock {\em Applied Optics}, 43(5):1024--1037, February 2004.

\bibitem{budil93}
K.~S. Budil, P.~Sali\`eres, Anne L’Huillier, T.~Ditmire, and M.~D. Perry.
\newblock Influence of ellipticity on harmonic generation.
\newblock {\em Physical Review A}, 48(5):R3437--R3440, November 1993.

\bibitem{soifer10}
H.~Soifer, P.~Botheron, D.~Shafir, A.~Diner, O.~Raz, B.~Bruner, Y.~Mairesse,
  B.~Pons, and N.~Dudovich.
\newblock Near-{Threshold} {High}-{Order} {Harmonic} {Spectroscopy} with
  {Aligned} {Molecules}.
\newblock {\em Physical Review Letters}, 105(14), September 2010.

\bibitem{fleischer14}
Avner Fleischer, Ofer Kfir, Tzvi Diskin, Pavel Sidorenko, and Oren Cohen.
\newblock Spin angular momentum and tunable polarization in high-harmonic
  generation.
\newblock {\em Nature Photonics}, 8(7):543--549, July 2014.

\bibitem{corkum93}
P.~B. Corkum.
\newblock Plasma perspective on strong field multiphoton ionization.
\newblock {\em Physical Review Letters}, 71(13):1994--1997, 1993.

\bibitem{paul06}
A.~Paul, E.A. Gibson, X.~Zhang, A.~Lytle, T.~Popmintchev, X.~Zhou, M.M.
  Murnane, I.P. Christov, and H.C. Kapteyn.
\newblock Phase-{Matching} {Techniques} for {Coherent} {Soft} {X}-{Ray}
  {Generation}.
\newblock {\em IEEE Journal of Quantum Electronics}, 42(1):14--26, January
  2006.

\bibitem{kazamias11}
S.~Kazamias, S.~Daboussi, O.~Guilbaud, K.~Cassou, D.~Ros, B.~Cros, and
  G.~Maynard.
\newblock Pressure-induced phase matching in high-order harmonic generation.
\newblock {\em Physical Review A}, 83(6), June 2011.

\bibitem{boguslavskiy12}
A.~E. Boguslavskiy, J.~Mikosch, A.~Gijsbertsen, M.~Spanner, S.~Patchkovskii,
  N.~Gador, M.~J.~J. Vrakking, and A.~Stolow.
\newblock The {Multielectron} {Ionization} {Dynamics} {Underlying} {Attosecond}
  {Strong}-{Field} {Spectroscopies}.
\newblock {\em Science}, 335(6074):1336--1340, March 2012.

\bibitem{sann16}
H.~Sann, T.~Havermeier, C.~Müller, H.-K. Kim, F.~Trinter, M.~Waitz,
  J.~Voigtsberger, F.~Sturm, T.~Bauer, R.~Wallauer, D.~Schneider, M.~Weller,
  C.~Goihl, J.~Tross, K.~Cole, J.~Wu, M. S. Sch\"offler,
  H.~Schmidt-B\"ocking, T.~Jahnke, M.~Simon, and R.~D\"orner.
\newblock Imaging the {Temporal} {Evolution} of {Molecular} {Orbitals} during
  {Ultrafast} {Dissociation}.
\newblock {\em Physical Review Letters}, 117(24):243002, December 2016.

\bibitem{fehre19}
K.~Fehre, S.~Eckart, M.~Kunitski, M.~Pitzer, S.~Zeller, C.~Janke, D.~Trabert,
  J.~Rist, M.~Weller, A.~Hartung, L.~Ph~H. Schmidt, T.~Jahnke, R.~Berger,
  R.~D\"orner, and M.~S. Sch\"offler.
\newblock Enantioselective fragmentation of an achiral molecule in a strong
  laser field.
\newblock {\em Science Advances}, 5(3):eaau7923, March 2019.

\bibitem{vos18}
J.~Vos, L.~Cattaneo, S.~Patchkovskii, T.~Zimmermann, C.~Cirelli, M.~Lucchini,
  A.~Kheifets, A.~S. Landsman, and U.~Keller.
\newblock Orientation-dependent stereo {Wigner} time delay and electron
  localization in a small molecule.
\newblock {\em Science}, 360(6395):1326--1330, June 2018.

\bibitem{powis14}
Ivan Powis.
\newblock Communication: {The} influence of vibrational parity in chiral
  photoionization dynamics.
\newblock {\em The Journal of Chemical Physics}, 140(11):111103, March 2014.

\bibitem{lucchese12}
Robert~R. Lucchese and Albert Stolow.
\newblock Molecular-frame photoelectron angular distributions.
\newblock 45(19):190201.

\bibitem{tia17}
Maurice Tia, Martin Pitzer, Gregor Kastirke, Janine Gatzke, Hong-Keun Kim,
  Florian Trinter, Jonas Rist, Alexander Hartung, Daniel Trabert, Juliane
  Siebert, Kevin Henrichs, Jasper Becht, Stefan Zeller, Helena Gassert, Florian
  Wiegandt, Robert Wallauer, Andreas Kuhlins, Carl Schober, Tobias Bauer,
  Natascha Wechselberger, Phillip Burzynski, Jonathan Neff, Miriam Weller,
  Daniel Metz, Max Kircher, Markus Waitz, Joshua~B. Williams, Lothar Ph.~H.
  Schmidt, Anne~D. Müller, Andr\'e Knie, Andreas Hans, Ltaief Ben~Ltaief, Arno
  Ehresmann, Robert Berger, Hironobu Fukuzawa, Kiyoshi Ueda, Horst
  Schmidt-B\"ocking, Reinhard D\"orner, Till Jahnke, Philipp~V. Demekhin, and
  Markus Sch\"offler.
\newblock Observation of {Enhanced} {Chiral} {Asymmetries} in the
  {Inner}-{Shell} {Photoionization} of {Uniaxially} {Oriented} {Methyloxirane}
  {Enantiomers}.
\newblock {\em The Journal of Physical Chemistry Letters}, 8(13):2780--2786,
  July 2017.

\bibitem{rozen19}
S.~Rozen, A.~Comby, E.~Bloch, S.~Beauvarlet, D.~Descamps, B.~Fabre, S.~Petit,
  V.~Blanchet, B.~Pons, N.~Dudovich, and Y.~Mairesse.
\newblock Controlling {Subcycle} {Optical} {Chirality} in the {Photoionization}
  of {Chiral} {Molecules}.
\newblock {\em Physical Review X}, 9(3):031004, July 2019.

\bibitem{lu_ground_1995}
Y.~Lu, Y.~Morioka, T.~Matsui, T.~Tanaka, H.~Yoshii, R.~I. Hall, T.~Hayaishi,
  and K.~Ito.
\newblock Ground and excited states of xe+2 observed by high resolution
  threshold photoelectron spectroscopy of xe2.
\newblock 102(4):1553--1560.

\bibitem{comby16}
Antoine Comby, Samuel Beaulieu, Martial Boggio-Pasqua, Dominique Descamps,
  Francois L\'egar\'e, Laurent Nahon, St\'ephane Petit, Bernard Pons, Baptiste
  Fabre, Yann Mairesse, and Val\'erie Blanchet.
\newblock Relaxation {Dynamics} in {Photoexcited} {Chiral} {Molecules}
  {Studied} by {Time}-{Resolved} {Photoelectron} {Circular} {Dichroism}:
  {Toward} {Chiral} {Femtochemistry}.
\newblock {\em The Journal of Physical Chemistry Letters}, 7(22):4514--4519,
  November 2016.

\end{thebibliography}
\end{document}